\documentstyle[12pt]{article}
\renewcommand{\thefootnote}{\fnsymbol{footnote}}
\begin{document}
\begin{flushright}
Columbia preprint CU--TP--693
\end{flushright}
\vspace*{1cm}
\setcounter{footnote}{1}
\begin{center}
{\Large\bf Relativistic Hydrodynamics for Heavy--Ion Collisions}\\
{\Large\bf II. Compression of Nuclear Matter and the Phase Transition to
the Quark--Gluon Plasma\footnote{This work was supported by the
Director, Office of Energy
Research, Division of Nuclear Physics of the Office of High Energy and Nuclear
Physics of the U.S. Department of Energy under Contract No.\
DE-FG-02-93ER-40764.}}
\\[1cm]
Dirk H.\ Rischke\footnote{Partially supported by the
Alexander von Humboldt--Stiftung under
the Feodor--Lynen program.} \\ ~~ \\
{\small Physics Department, Pupin
Physics Laboratories, Columbia University} \\
{\small 538 W 120th Street, New
York, NY 10027, U.S.A.} \\ ~~ \\
Yar\i\c{s} P\"urs\"un, Joachim A.\ Maruhn \\ ~~ \\
{\small Institut f\"ur Theoretische Physik der J.W.\ Goethe--Universit\"at} \\
{\small Robert--Mayer--Str.\ 10, D--60054 Frankfurt/M., Germany}
\\ ~~ \\ ~~ \\
{\large April 1995}
\\[1cm]
\end{center}
\begin{abstract}
We investigate the compression of nuclear matter in
relativistic hydrodynamics. Nuclear matter is described by a
$\sigma-\omega$--type model for the hadron matter phase and by
the MIT bag model for the quark--gluon plasma, with a first order phase
transition between both phases. In the presence of phase transitions,
hydrodynamical solutions change qualitatively, for instance,
one-dimensional stationary compression is no longer accomplished
by a single shock but via a sequence of shock and compressional simple waves.
We construct the analytical solution to the ``slab-on-slab'' collision
problem over a range of incident velocities. The performance of numerical
algorithms to solve relativistic hydrodynamics is then investigated
for this particular test case.
Consequences for the early compressional stage in heavy--ion collisions are
pointed out.
\end{abstract}
\renewcommand{\thefootnote}{\arabic{footnote}}
\setcounter{footnote}{0}
\newpage
\section{Introduction}

To investigate heavy--ion collision dynamics in realistic, i.e.,
3+1--dimensional, situations by means of hydrodynamics requires
numerical schemes to solve the hydrodynamical equations of motion.
These schemes should reproduce analytical solutions, as far as such exist
at all.
In a previous paper \cite{test1} we have presented two algorithms
for ideal relativistic hydrodynamics, the SHASTA, a
flux--corrected transport algorithm \cite{SHASTA}, and the
relativistic HLLE, a Godunov--type algorithm \cite{schneider}.
We have investigated their performance
for the expansion of matter into the vacuum,
where they are confronted with two problems generical for simulations of
heavy--ion collisions. These are the presence of vacuum
itself and the qualitative change in the type of the hydrodynamical
expansion solution if phase transitions occur in the equation of state (EoS).

As was explained in detail in \cite{test1,bugaev},
matter that undergoes a first order phase transition may exhibit
thermodynamically anomalous behaviour in a certain range
of independent thermodynamic variables,
signalled by a change of sign of the quantity
\begin{equation} \label{Sigma}
\Sigma \equiv \left. \frac{\partial^2 p}{\partial \epsilon^2} \right|_{\sigma}
 + 2\, c_s^2~\frac{1-c_s^2}{\epsilon +p}~,
\end{equation}
where $c_s^2 \equiv \partial p/ \partial \epsilon |_{\sigma}$ is the
velocity of sound, $\epsilon, p,$ and $\sigma$ are the energy density,
pressure, and specific entropy, respectively.
For thermodynamically normal (TN) matter, $\Sigma >0$, for
so-called thermodynamically anomalous (TA) matter $\Sigma < 0$
\cite{test1,bugaev}.

As was shown in \cite{test1,bugaev,olli}, for the one--dimensional
expansion of TN matter a simple rarefaction wave is the stable
hydrodynamical solution, while for TA matter such a wave is unstable with
respect to formation of a rarefaction shock wave.
Moreover, an initial discontinuity is bound to decay into a simple wave in TN
matter, but cannot do so if matter is TA (or even if
$\Sigma$ only vanishes instead of becoming negative).
Thus, rarefaction discontinuities form
the stable hydrodynamical solution in the expansion of TA matter.

On the other hand, for the compression of TA matter a compressional simple
wave forms the stable hydrodynamic solution, but it is
unstable in TN matter with respect to formation of a compressional
shock wave. Analogously, a compressional shock discontinuity is bound to
decay into a compressional simple wave in TA matter,
while it cannot do so in TN matter.
Thus, compression shocks are stable in the compression of TN matter.

As outlined in \cite{test1}, in realistic cases the EoS has
both TN and TA regions. Consequently,
the hydrodynamical solution for the one--dimensional
expansion is more complicated, involving
a sequence of simple waves, regions of constant flow, and discontinuities.
It was shown in \cite{bugaev} that the same holds for the
compression. The corresponding hydrodynamical solution was explicitly
constructed in the case of a one--dimensional ``slab-on-slab''
collision. (For TN matter this test problem and the corresponding performance
of the SHASTA and relativistic HLLE algorithms was studied
in \cite{schneider}.)
The nuclear matter equations of state used in \cite{bugaev} featured
a first order phase transition between the quark--gluon plasma (QGP),
described by the MIT bag model, and hadronic matter, described by
phenomenological equations of state \cite{walecka,zim}.

In this work we first construct a nuclear matter EoS similar to that
of Ref.\ \cite{bugaev}. We also use the MIT bag EoS \cite{MIT} for the
QGP phase, but for the hadronic phase, we employ a version of the
$\sigma-\omega$--model \cite{migdhr} (plus massive thermal pions)
which features more realistic values for the ground state incompressibility
and the effective nucleon mass than the original version proposed
by Walecka \cite{walecka}.
For this EoS, we construct analytically the hydrodynamical compression
solution as described in \cite{bugaev}. With a tabulated version of
the EoS, we then test the ability of the numerical algorithms to
reproduce this solution. This
does not only yield insight into the usefulness of these algorithms for
applications in heavy--ion collisions and dynamical studies
of QGP formation. In a sense it also represents an independent
``numerical'' proof for the theoretical arguments of
Ref.\ \cite{bugaev} about the form of the hydrodynamical solution
for compression of TA matter.

This paper is organized as follows. In Section 2 we present our nuclear matter
EoS. In Section 3 we analytically construct the hydrodynamic solution for
a one--dimensional ``collision'' of semi-infinite slabs.
Section 4 comprises our numerical results.
In Section 5 we comment on one--dimensional collisions of
finite nuclei and draw conclusions for numerical simulations
of heavy--ion collisions. Section 6 concludes this work with a summary
of our results. An Appendix contains
results for modifications of the standard algorithms used
throughout Ref.\ \cite{test1} and this work.
We use natural units $\hbar=c=k_B=1$.

\section{The nuclear matter EoS}

\subsection{Hadron matter}

For hadronic matter we use a modification of the original
$\sigma-\omega$--model \cite{walecka}, which
was presented in Ref.\ \cite{migdhr}. For the original
$\sigma-\omega$--model, the EoS, i.e., the pressure $p$ as a
function of the independent thermodynamical
variables temperature $T$ and baryochemical potential $\mu$,
can be rigorously derived from the $\sigma-\omega$--Lagrangian
employing the mean--field (or Hartree, or one--loop) approximation of
quantum many--body theory at finite temperature and density
\cite{fetter}, see for instance \cite{dhrwg} for an explicit
derivation.
A more general class of thermodynamically consistent,
phenomenological equations of state for (non--strange)
interacting nucleonic matter is defined by \cite{gori}
\begin{eqnarray} \label{phad}
p_{had}(T,\mu) & = & p_N(T,\nu;M^*) + p_N(T,-\nu;M^*) + \sum_i p_i(T;m_i)\\
         & + & n\, {\cal V}(n) - \int_0^n {\cal V}(n')\, {\rm d}n'
               - \rho_s\, {\cal S}(\rho_s) + \int_0^{\rho_s}
               {\cal S}(\rho_s')\, {\rm d}\rho_s'~. \nonumber
\end{eqnarray}
Here,
\begin{equation} \label{pn}
p_N(T,\nu;M^*) = T\, \frac{g_N}{(2\pi)^3} \int {\rm d}^3 {\bf k}~
       \ln \left[ 1+\exp \left\{-\left( E_k^* -\nu \right)/T \right\} \right]
\end{equation}
is the pressure of an ideal gas of nucleons (spin--isospin degeneracy
$g_N=4$) moving in the scalar potential ${\cal S}$ and the
vector potential ${\cal V}$ \footnote{ More accurately, ${\cal V}$ is
the zeroth component of the vector potential ${\cal V}^{\mu}$, the
spatial components of which vanish in a homogeneous, isotropic
system.}. These potentials generate an {\em effective\/} nucleon mass
\begin{equation}
M~~ \longrightarrow~~ M^* \equiv M - {\cal S}(\rho_s)~,
\end{equation}
where $M= 938$ MeV is the nucleon mass in the vacuum, and also shift the
one--particle energy levels
\begin{equation}
E_k \equiv \sqrt{{\bf k}^2 + M^2}~~\longrightarrow~~ E_k^* +{\cal V}(n) \equiv
\sqrt{{\bf k}^2 + \left(M^*\right)^2} + {\cal V}(n)~.
\end{equation}
The vector potential is conveniently absorbed in the {\em effective\/}
chemical potential
\begin{equation}
\nu \equiv \mu - {\cal V}(n)~,
\end{equation}
giving rise to the interpretation of (\ref{pn})
as the pressure of an ideal gas of {\em quasi-particles\/} with mass
$M^*$ and chemical potential $\nu$.
Furthermore,
\begin{equation}
p_i(T;m_i) = -T \,\frac{g_i}{(2\pi)^3} \int
 {\rm d}^3 {\bf q}~\ln \left[ 1-\exp \left\{ - \sqrt{{\bf q}^2 +m_i^2}/T
\right\} \right]
\end{equation}
is the pressure of an ideal gas of mesons with degeneracy $g_i$
and mass $m_i$. In the following, we will only include pions
($g_{\pi}=3,\, m_{\pi}=138$ MeV), since
they are the lightest and thus most abundant mesons.
(To be consistent, one should also include
thermal contributions for the $\sigma$ and $\omega$ meson \cite{dhrwg}.
However, since these mesons are much heavier than the pion, one can
neglect their contribution for the present applications.)
$n$ is the (net) baryon density,
\begin{eqnarray} \label{n}
n(T,\mu) & \equiv & \left. \frac{\partial p}{\partial \mu} \right|_T \\
 &  = &\frac{g_N}{(2\pi)^3} \int {\rm d}^3 {\bf k}~\left[
  \frac{1}{e^{(E_k^* - \nu)/T}+1} - \frac{1}{e^{(E_k^* + \nu)/T}+1} \right]~,
\nonumber
\end{eqnarray}
and
\begin{equation}
\rho_s(T,\mu) \equiv \frac{g_N}{(2\pi)^3} \int {\rm d}^3
{\bf k}~\frac{M^*}{E_k^*}~ \left[ \frac{1}{e^{(E_k^* - \nu )/T}+1}
              + \frac{1}{e^{(E_k^* + \nu )/T}+1} \right]
\end{equation}
is the scalar density of nucleons. It is assumed that the
vector potential (which transforms like the zeroth component of a vector)
depends {\em only\/} on the (net) baryon density (which also transforms
like the zeroth component of a vector, namely the (net) baryon current
$N^{\mu}$), while the scalar potential (which is a Lorentz--scalar)
depends only on the (Lorentz--) scalar density $\rho_s$.

Once the pressure is known, the entropy and energy density can be obtained
from the thermodynamical relations,
\begin{equation} \label{s}
s = \left. \frac{\partial p}{\partial T} \right|_{\mu}~,~~\epsilon =
Ts + \mu n - p~.
\end{equation}
One now has to specify the potentials ${\cal V,S}$. The choice
\begin{equation}
{\cal V}(n) = C_V^2\, n~,~~{\cal S}(\rho_s)= C_S^2\, \rho_s
\end{equation}
reproduces the original $\sigma-\omega$--model \cite{dhrwg}.
When fitting the two parameters $C_V^2,\, C_S^2$ to reproduce
the experimental values for the ground state binding energy of infinite nuclear
matter, $B_0 = 16$ MeV, and the ground state density, $n_0 = 0.15891\,
{\rm fm}^{-3}$ \cite{friedman}, the effective mass in the nuclear ground state
is $M^*_0 \simeq 0.543\, M$ which is too small
(the experimental value ranges around $0.7\, M$
\cite{effmass}), and the incompressibility is $K_0 \equiv 9\,
{\rm d}p/{\rm d}n|_{n_0} \simeq 553$ MeV which is too large by about
a factor of two \cite{comp}.
To adjust this shortcoming of the model, it was suggested to introduce
self-interaction terms, $\sim \sigma^3,\, \sigma^4$,
for the scalar $\sigma$ meson field in the
$\sigma -\omega$--Lagrangian \cite{nonlin}.
This nonlinear $\sigma-\omega$--model has
two additional parameters which enable one to also independently adjust
$M^*_0$ and $K_0$ to the experimentally observed values.

Alternatively, in Ref.\ \cite{migdhr} it was suggested to use
\begin{equation} \label{choice}
{\cal V}(n) = C_V^2\, n - C_d^2\, n^{1/3}~,~~{\cal S}(\rho_s) = C_S^2\,
\rho_s~.
\end{equation}
Although this choice introduces only {\em one\/} additional parameter, fitting
the ground state incompressibility gives {\em simultaneously\/}
reasonable values for the effective mass, for details see \cite{migdhr}.
In the following we use the parameters $C_V^2= 238.08\, {\rm GeV}^{-2},\,
C_S^2 = 296.05\, {\rm GeV}^{-2},\, C_d^2 = 0.183$, which leads
to $M^*_0 = 0.635\, M,\, K_0 = 300\, {\rm MeV}$.

\subsection{Quark--Gluon Plasma}

For the QGP EoS we employ the standard MIT bag model \cite{MIT}
for massless, non-interacting gluons and $u,\, d$ quarks, i.e.,
\begin{equation} \label{pQGP}
p_{QGP}(T,\mu) = \frac{37 \pi^2}{90}~T^4 + \frac{1}{9}~\mu^2 T^2 +
\frac{1}{162 \pi^2}~\mu^4 -B~,
\end{equation}
and other thermodynamical quantities follow again from the
relations (\ref{n}, \ref{s}).
Note that $p$ does not depend explicitly on $n$ for this EoS,
\begin{equation} \label{pQGP2}
p = \frac{1}{3} (\epsilon - 4B)~.
\end{equation}
For the bag constant, we use the value
$B=(235\, {\rm MeV})^{4}$ throughout this work. This value
results in a phase transition temperature of $T_c \simeq 169$ MeV at
vanishing baryon density (see below), which is roughly in agreement with
what lattice QCD simulations predict \cite{tc}.

\subsection{Gibbs Phase Equilibrium and Thermodynamical Phase Diagrams}

The QGP EoS (\ref{pQGP}) is matched to the hadronic EoS
(\ref{phad}) with (\ref{choice})
via Gibbs' conditions for (mechanical, thermal, and chemical)
phase equilibrium,
\begin{equation} \label{Gibbs}
p_{had} = p_{QGP}~,~~T_{had} = T_{QGP}~,~~\mu_{had} = \mu_{QGP}~,
\end{equation}
which leads to a phase boundary curve $T^*(\mu^*)$ in the
$T-\mu$ plane defined by the implicit equation $p_{had}(T^*,\mu^*) =
p_{QGP}(T^*,\mu^*)$, see Fig.\ 1 (a).
Along this curve, one can calculate the phase boundary values
for other thermodynamical variables as a function of $T^*$.
Fig.\ 1 (b) shows the phase boundaries $n_{H,Q}(T^*)$ in the
$T-n/n_0$ phase diagram, Fig.\ 1 (c) the phase boundaries
$\epsilon_{H,Q}(T^*)$ in the $\epsilon/\epsilon_0 -T$ diagram, and
Fig.\ 1 (d) $\epsilon_H(n_H(T^*))$ and $\epsilon_Q(n_Q(T^*))$ in the
$\epsilon/\epsilon_0 - n/n_0$ diagram, respectively
($\epsilon_0=(M-B_0)n_0$ is the ground state energy density).
The phase transition constructed via (\ref{Gibbs}) is of
first order, leading to a mixed phase of QGP and hadron matter
and to a latent heat, as can be seen in Figs.\ 1 (b,c,d).
The $T$--axis of Fig.\ 1 (b) maps onto the (dotted) curve
$\epsilon(T,n=0)$ in Fig.\ 1 (c). Correspondingly,
the $\epsilon/\epsilon_0$--axis in Fig.\ 1 (c) maps onto the (dotted) curve
$\epsilon(n,T=0)$ in Fig.\ 1 (d).
This zero--temperature energy density is finite for finite $n$
due to the Fermi (and, in the hadronic phase, interaction) energy of the
fermions in the system (nucleons and quarks, respectively).
This curve represents the {\em minimum\/} energy density possible for a
given baryon density.

As described in \cite{test1}, the pressure as a function of energy density
and baryon density, $p(\epsilon,n)$, is required for solving
the hydrodynamical equations numerically. It is convenient to tabulate
this function, since calculating the pressure in the
hadron phase for given $\epsilon, \, n$
requires a triple root search (a double root search to
find $T,\, \mu$ for given $\epsilon,\, n$, while simultaneously solving a
fixed-point equation for the effective nucleon mass). The computational
effort would be prohibitive if one tries to perform this ``on-line''
for each hydrodynamic cell in each time step of the hydrodynamical
transport algorithm.

Therefore, we discretize the $\epsilon-n$ plane of Fig.\ 1 (d) in the
range $0 \leq \epsilon/\epsilon_0 \leq 20, \, 0 \leq n/n_0 < 12$
with equidistant grid spacing $\Delta \epsilon = \epsilon_0/10,
\, \Delta n = n_0/20$ to form a $201 \times 240$ mesh. This
grid spacing proves to be sufficiently small to use the resulting
table also for analytical calculations, see Section 4.
The phase boundaries $\epsilon_{H,Q}$,
as well as the $\epsilon(n,T=0)$--curve are mapped onto the
240 grid points of the $n/n_0$--axis ($\epsilon_{H,Q}$ is taken to be
zero for $n$--values exceeding those where the phase boundaries meet the
$\epsilon(n,T=0)$--curve).

Now values for the pressure are assigned to the mesh points. In
the hadronic phase (defined by $\epsilon(n,T=0) \leq \epsilon
 \leq \epsilon_H(n)$ for given $n$) this is done using eq.\
(\ref{phad}) and the mentioned triple root search, and
in the QGP phase (defined by ${\rm max}
\{\epsilon_Q(n),\epsilon(n,T=0)\}
\leq \epsilon \leq 20\, \epsilon_0$ for given $n$) eq.\ (\ref{pQGP2})
is employed. In the mixed phase
(${\rm max} \{\epsilon_H(n),\epsilon(n,T=0)\} \leq \epsilon \leq
\epsilon_Q(n)$ for given $n$) the pressure is calculated as follows.
For given $T^*$, the values of energy and baryon density read
\begin{eqnarray} \label{e}
\epsilon & = & \lambda_Q \epsilon_Q(T^*) + (1-\lambda_Q) \epsilon_H(T^*)~, \\
n & = & \lambda_Q n_Q(T^*) + (1-\lambda_Q) n_H(T^*)~, \label{n2}
\end{eqnarray}
where $\lambda_Q$ is the fraction of volume the QGP phase occupies in
the mixed phase.
Vice versa, for given $\epsilon,\, n$ these equations yield values for
$\lambda_Q,\, T^*$. Eliminating $\lambda_Q$ one obtains a single equation
for $T^*$ as a function of $\epsilon,\, n$. This is solved numerically
using the values $\epsilon_{H,Q}(T^*),\, n_{H,Q}(T^*)$ of Figs.\ 1 (b,c).
Once $T^*$ is known, $\mu^*$ follows from Fig.\ 1 (a), which
consequently yields the pressure
from Gibbs' phase equilibrium conditions (\ref{Gibbs}).

Similarly, every other thermodynamic quantity can be calculated
as a function of $\epsilon$ and $n$. Of particular importance are
temperature, baryo-chemical potential, and entropy density.
Once the first two are known, the entropy density is obtained
from the second equation (\ref{s}).
In the hadronic phase, $T$ and $\mu$ emerge naturally from the
triple root search, while in the mixed phase $T^*$ is obtained
as described above and $\mu^*$ then follows from Fig.\ 1 (a).
In the QGP, we distinguish two cases. For $n=0$ we also have $\mu=0$,
and then we obtain from (\ref{pQGP}, \ref{pQGP2}) the simple formula
$T= [30(\epsilon-B)/37 \pi^2]^{1/4}$.
For finite $\mu$ we eliminate $T$ from the equation of $\epsilon$ using
the equation for $n$. This results in a sixth order equation in
$\mu$,
\begin{equation} \label{sixt}
0 = \frac{4}{1215 \pi^2}~\mu^6 - \frac{4}{15}~n \mu^3 + (\epsilon -B) \mu^2
- \frac{999 \pi^2}{40}~n^2~,
\end{equation}
which has to be solved numerically. After that, $T = [9n/2\mu -
\mu^2/9\pi^2]^{1/2}$ (which follows from the equation for $n$).

In Fig.\ 2 we show (a) pressure, (b) temperature, (c) baryo-chemical
potential, and (d) entropy density as functions of $\epsilon$ and $n$, as
they emerge from the above described procedure. Note in Fig.\ 2 (a)
that on account of (\ref{pQGP2})
the pressure is independent of $n$ in the QGP phase.
Also, in the mixed phase for small $n$
the pressure is (almost) constant as a function of either $\epsilon$ or $n$.
The thermodynamical identity
\begin{equation} \label{cs2}
c_s^2 \equiv \left. \frac{\partial p}{\partial \epsilon} \right|_{\sigma}
= \left. \frac{\partial p}{\partial \epsilon} \right|_n + \frac{n}{\epsilon
+p}\left. \frac{\partial p}{\partial n} \right|_{\epsilon}
\end{equation}
then implies that the velocity of sound is very small in this region.
For $n=0$, the pressure (as well as the temperature, see Fig.\ 2 (b))
is constant in the mixed phase, and the velocity of sound vanishes
identically (cf.\ the EoS for TA matter in \cite{test1}).

In contrast to the diagrams for the thermodynamical variables $p,T,\mu$
which enter Gibbs' phase equilibrium conditions (\ref{Gibbs}),
there is no obvious
sign for the mixed phase (such as edges due to the phase boundaries)
in the entropy density.
The reason is that this quantity is a linear interpolation
between $s_H$ and $s_Q$ (similar to (\ref{e}, \ref{n2})) in this phase.

For the hydrodynamical simulations intermediate values of
thermodynamic quantities are calculated by two--dimensional linear
interpolation. If $\epsilon> {\rm max} \{ 20\, \epsilon_0, \epsilon(n,T=0) \}$
for given $n$, the QGP EoS (\ref{pQGP2})
is used directly to calculate the pressure, and temperature, chemical
potential, and entropy density are obtained as described above.

\section{Compression of nuclear matter}

In this section we construct the analytical solution to
the one--dimensional compression of nuclear matter with the EoS presented
in Section 2. The initial condition of the ``slab-on-slab'' collision
problem is
\begin{eqnarray}
\epsilon(x,0) & = & \epsilon_0~~,~~~-\infty < x < \infty~,\\
n(x,0) & = & n_0~~,~~~-\infty < x < \infty~,\\
v(x,0) & = & \left\{ \begin{array}{ll}
                   v_{CM}~, & -\infty < x \leq 0~,\\
                   -v_{CM}~, & 0 < x < \infty~.
                 \end{array}    \right.
\end{eqnarray}
In a real heavy--ion collision (see also Section 5), $v_{CM}$ corresponds to
the velocity of the nuclei in the ``equal--velocity frame''. For identical
nuclei, this is equivalent to the center-of-momentum (CM) frame.
In the subsequent time evolution, momentum conservation will
force matter to stop at $x=0$ and
energy conservation to convert its kinetic energy into internal energy, leading
to compression waves travelling symmetrically away from the origin, leaving
in their wake compressed and heated matter at rest.
For small $v_{CM}$, the energy and baryon number densities obtained in
the compressed state correspond to hadronic, i.e., TN matter.
Thus, compression proceeds via single shock waves.
Energy, momentum, and (net) baryon number conservation across the
(space--like) shock discontinuity read in the rest frame of the shock
(quantities with subscript ``0'' correspond to the uncompressed state)
\begin{equation} \label{cons}
T^{01} = T^{01}_0~,~~T^{11}=T^{11}_0~,~~N^1=N^1_0~,
\end{equation}
where $T^{\mu \nu}=(\epsilon + p)u^{\mu}u^{\nu}-pg^{\mu \nu}$ is
the energy momentum tensor for an ideal fluid, and $N^{\mu}=nu^{\mu}$ is
the (net) baryon number current.
The final compressed
state $(\epsilon, n)$ obeys the {\em Taub equation\/} \cite{taub}
\begin{equation} \label{Teq}
wX - w_0 X_0 - (p-p_0)(X+X_0) = 0~,
\end{equation}
where $w=\epsilon+p$ is the enthalpy density and $X=w/n^2$ the generalized
volume\footnote{Note that this definition differs from the one given
in \cite{test1} for the (net) baryon-free case.}.
The Taub equation defines the so-called {\em Taub adiabat\/} in the
$p-X$ plane. It is the set of all possible final states which are in
accordance with energy, momentum, and baryon number conservation across
a single shock discontinuity.
A particular final state is fixed by specifying
the involved velocities. For instance, in our case the velocity $v_{CM}$ of
uncompressed matter in the rest frame of compressed matter is given.
On the other hand, in the rest frame of the shock, uncompressed matter
flows into the shock with velocity $v_0$ and compressed matter emerges
with velocity $v$. These velocities obey \cite{taub}
\begin{equation} \label{v}
v_0^2= \frac{(p-p_0)(\epsilon + p_0)}{(\epsilon-\epsilon_0)(\epsilon_0+p)}~,~~
v^2= \frac{(p-p_0)(\epsilon_0 + p)}{(\epsilon-\epsilon_0)(\epsilon+p_0)}~.
\end{equation}
Obviously, $v_{CM} = |v_0-v|/(1-v_0v)$ and thus
\begin{equation} \label{gamma}
\gamma_{CM}^2 = \frac{(\epsilon_0 +p)(\epsilon+p_0)}{w\, w_0} =
\left[ \frac{(\epsilon+p_0)\, n_0}{w_0\, n}
\right]^2 \equiv \left( \frac{\epsilon/n}{\epsilon_0/n_0} \right)^2~,
\end{equation}
since $p_0 \equiv 0$ in the ground state of nuclear matter. In deriving
this equation we have made use of the Taub equation (\ref{Teq}). Given
$v_{CM}$, this equation fixes $\epsilon/n$, or with the Taub equation,
$\epsilon$ {\em and\/} $n$ in the final compressed state.

Fig.\ 3 (a) shows the Taub adiabat with the center
$(X_0,p_0) = (\epsilon_0/n_0^2,0)$ for the EoS of Section 2.
The different parts belonging to final states in the hadron, mixed, and
QGP phase are marked correspondingly.
One observes a region of final states (A--C)
where the chord connecting them with the center intersects the Taub adiabat.
For these final states a single compression shock is not the
hydrodynamically stable solution \cite{bugaev,gori,LL6}.

To see this and to determine the stable solution one first notes
that point A is a Chapman-Jouguet (CJ) point of the Taub adiabat. At this
point, the specific entropy $\sigma \equiv s/n$ assumes a (local) maximum and
the slope of the Taub adiabat and the Poisson adiabat (the curve of
constant specific entropy) are identical \cite{LL6}
\begin{equation} \label{ident}
\left. \frac{\partial X}{\partial p} \right|_{\sigma} \equiv
\left. \frac{\partial X}{\partial p} \right|_{TA}~.
\end{equation}
Furthermore, above the CJ point the chord connecting a final state
$(X,p)$ with the center has a smaller slope than the Taub
adiabat at $(X,p)$, i.e.,
\begin{equation} \label{ineq}
\left. \frac{\partial X}{\partial p} \right|_{TA} \leq
\frac{X-X_0}{p-p_0}~~,~~~p>p_{CJ}~,
\end{equation}
with the equality holding at the CJ point.
Subtracting this equality from (\ref{ineq}), we obtain
\begin{equation}
\frac{ \partial X/\partial p|_{TA,\, p} - \partial X/\partial p |_{\sigma,
\, p_{CJ}} }{p-p_{CJ}} \leq 0~~,~~~p>p_{CJ}~.
\end{equation}
In the limit $p \rightarrow p_{CJ}$ we conclude using (\ref{ident}):
\begin{equation}
\left. \frac{\partial^2 X}{\partial p^2} \right|_{\sigma,\, p_{CJ}}
 \equiv \tilde{\Sigma} \leq 0 ~~{\rm for}~~p > p_{CJ}~.
\end{equation}
{}From thermodynamical identities, $\tilde{\Sigma} \equiv \Sigma/n^2 c_s^6$,
and therefore, {\em above the CJ point matter is TA}. This implies that,
once the CJ point is reached via a single compression shock, further
compression can be achieved only through a {\em compressional simple wave}.
The hydrodynamical solution will therefore consist of a single compression
shock to the CJ point and, attached to it, a compressional simple wave
(see Fig.\ 5 below). Since matter emerges from the shock with the velocity
of sound (since the compressed state corresponds to the CJ
point \cite{LL6}), and since for a simple wave the matter velocity
relative to the wave profile is also the velocity of sound,
the simple wave profile does not move relative to the shock, and they
will remain attached to each other in the stationary state.
Since the specific entropy is constant for continuous solutions of
ideal hydrodynamics, final states of compressed matter lie on the
Poisson adiabat through the CJ point, rather than on the Taub adiabat,
see Fig.\ 3 (b).

Compression of matter can proceed through this configuration only until
the point of inflection, $\partial^2 X/\partial p^2|_{\sigma} = 0$,
on the Poisson adiabat is reached (point B in Fig.\ 3 (b); this is
usually the point where the Poisson adiabat leaves the mixed phase and enters
the QGP phase). Since matter becomes TN at this point,
further compression is achieved by another compression
shock attached to the head of the compressional simple wave (see Fig.\ 6
below).
This shock is also described by the Taub equation (\ref{Teq}), however,
$X_0$, $p_0$, and $w_0$ have to replaced by the corresponding quantities
$\tilde{X}$, $\tilde{p}$, and $\tilde{w}$
at the head of the compressional simple wave. The final compressed state
lies in the pure QGP phase and is determined as follows.
For a stationary configuration, the (baryon) flux
$N^1=n\gamma v$ at the head
of the compressional simple wave has to be identical to that through
the shock front. In the rest frame of the latter it is determined by
$ (N^1)^2 = - (p-\tilde{p})/(X-\tilde{X})$ (which follows from
the conservation laws (\ref{cons}); the interpretation is that the square
of the baryon flux through the shock equals the negative slope of the chord
connecting initial and final state of the shock). On the other hand,
the matter velocity at a fixed point of a simple wave profile is equal to the
velocity of sound, i.e., in the rest frame of that point the
baryon flux is $N^1 = \tilde{n} c_s/(1-c_s^2)^{1/2}$. Now consider
this point to be the head of the simple wave, or equivalently,
the position of the compression shock, if the configuration is stationary.
Since $\partial p/\partial X|_{\sigma} = - n^2 c_s^2/
(1-c_s^2)$ (which follows from thermodynamical identities),
we arrive at the condition
\begin{equation} \label{cond2}
\left. \frac{\partial p}{\partial X} \right|_{\sigma,\, \tilde{X}}
\equiv \frac{p-\tilde{p}}{X-\tilde{X}}
\end{equation}
for a stationary hydrodynamical configuration.
This means that the final compressed state $(X,p)$ on the Taub adiabat
with center $(\tilde{X},\tilde{p})$ is determined by the intersection
of this adiabat with the tangent to the Poisson adiabat at $(\tilde{X},
\tilde{p})$.

In order to reach higher compression through this shock,
one has to decrease the slope
(\ref{cond2}), i.e., the state $(\tilde{X},\tilde{p})$
has to ``move backwards'' along the Poisson adiabat towards the CJ
point A. The compressional wave will consequently become ``shorter''
and the compression shock stronger. This proceeds until the CJ point A
is reached. Then, the compressional simple wave vanishes and the shock
(O--A) merges with the shock (A--C).
Since the chord (A--C) has the same slope as the chord (O--A), the baryon flux
through the two shocks is identical and therefore, the single shock
(O--C) becomes the hydrodynamically stable, stationary configuration.
Above C, the single shock described by the Taub adiabat of Fig.\ 3 (a)
represents again the stable hydrodynamical solution.

In practice, one obtains part (B--C), the so-called {\em wave
adiabat\/} \cite{bugaev}, when calculating the Poisson
adiabat: for each $(\tilde{X},\tilde{p})$ one
simply solves (\ref{cond2}) simultaneously with
the Taub equation with center $(\tilde{X},\tilde{p})$. The set of all final
compressed states $(X,p)$ thus obtained was called {\em generalized
shock adiabat\/} in Ref.\ \cite{bugaev}.

Let us now complete the solution of the hydrodynamical problem.
As was the case for the expansion of semi--infinite matter
studied in \cite{test1},
the stationary hydrodynamical solution is of
similarity type, i.e., its profile is constant as a function of
$\zeta \equiv x/t$. Due to causality and the symmetry of the problem, it
suffices to consider the range $0\leq \zeta\leq 1$.
Below and at the CJ point,
compression proceeds through a single shock, travelling
to the right with velocity $v_{sh} \equiv -v$, where $v$ is given by
(\ref{v}). Given $v_{CM}$, the final state energy density $\epsilon$ and
baryon number density $n$ are
obtained as solutions of (\ref{Teq}) and (\ref{gamma}).
Other hydrodynamical variables can be inferred from the EoS and the definition
of $T^{\mu \nu}$ and $N^{\mu}$. Profiles of $T^{00}/\epsilon_0$ and
$T$ as functions of $\zeta$
are shown in Fig.\ 4 for $v_{CM} = 0.7$.

The CJ point itself has to be determined numerically by the requirement of
maximum specific entropy. For the above EoS, $p_{CJ} \simeq 1.941\, \epsilon_0,
\, X_{CJ} \simeq 0.422\, X_0$, ($\epsilon_{CJ} \simeq 6.971\, \epsilon_0,\,
n_{CJ} \simeq 4.596\, n_0$,) with a specific entropy $\sigma \simeq 3.545$.
For further purpose, let us denote the value of $v_{CM}$ required
to reach the CJ point by $v_{CJ}$. From (\ref{gamma}) we infer
$v_{CJ} \simeq 0.752$.

Above the CJ point and below point B,
compression proceeds through a shock to the CJ point
and an attached compressional simple wave. The shock moves with
$v_{sh} = [v_{sh}'-v_{CM}]/[1-v_{sh}'\, v_{CM}]$ where $v_{sh}'$ is the
shock velocity in the rest frame of incoming matter. The latter is identical
with $-v_0$, cf.\ (\ref{v}), i.e., $v_{sh}' \simeq 0.878$. Obviously,
matter emerges from the shock with velocity $v_{CJ}$ in the rest frame
of incoming matter, i.e., in the CM frame matter
emerges from the shock with velocity $[v_{CJ}-v_{CM}]/[1-v_{CJ} v_{CM}]$.
This is, of course, still negative above the CJ point,
since the subsequent compressional simple wave will further decelerate
matter.

Consider the compressional simple wave moving to the left (i.e., that at
$x>0$) in the rest frame of matter emerging from
the shock. Constancy of the Riemann invariant ${\cal R}_-$
for this wave \cite{test1} implies that the velocity
as a function of the energy density reads
\begin{equation} \label{vw}
v_{w}''(\epsilon_w) = \tanh \left\{ \int_{\epsilon_{CJ}}^{\epsilon_w}
{\rm d} \epsilon'~\frac{c_s(\epsilon')}{\epsilon' + p(\epsilon')}
\right\}~.
\end{equation}
The integral is calculated via a centered Riemann sum. The required
values of $p(\epsilon')$ and $\epsilon'$ are taken from Fig.\ 3 (b).
In that calculation it was also necessary to locally
determine the tangent to the Poisson adiabat (in order to construct the wave
adiabat (B--C)). The slope of the tangent is $\partial p/\partial X|_{\sigma}
\equiv -n^2 c_s^2/(1-c_s^2)$ and this relation can be easily inverted
to yield the local velocity of sound $c_s(\epsilon')$ required for
the integration in (\ref{vw}).

In the rest frame of incoming matter, the velocity of matter on the
simple wave is $v_w'(\epsilon_w) = [v_w''(\epsilon_w) + v_{CJ}]/[1+
v_w''(\epsilon_w)\, v_{CJ}]$, and consequently in the CM frame
$v_w(\epsilon_w) = [v_w'(\epsilon_w)-v_{CM}]/[1-v_w'(\epsilon_w)\, v_{CM}]$.
The complete simple wave profile is obtained by successively increasing
$\epsilon_w$ in (\ref{vw}) until $v_w$ becomes zero. Then, the simple wave
has completely decelerated matter in the central region.

The similarity variable $\zeta$ corresponding to a given $\epsilon_w$ is
determined via \cite{test1}
\begin{equation} \label{zeta}
\zeta = \frac{v_w(\epsilon_w) + c_s(\epsilon_w)}{1 + v_w(\epsilon_w)\,
c_s(\epsilon_w)}~.
\end{equation}
The final energy density $\epsilon$ where matter comes to rest determines
the position $\zeta \equiv c_s(\epsilon)$ of the head of the simple wave.
Profiles of $T^{00}/\epsilon_0$ and
$T$ as functions of $\zeta$
are shown in Fig.\ 5 for $v_{CM} = 0.8$.

Point B on the generalized shock adiabat is numerically determined as
$p_B \simeq 2.478\, \epsilon_0$, $ X_B \simeq 0.198\, X_0$ ($\epsilon_B
\simeq 18.271\, \epsilon_0,\, n_B \simeq 10.236\, n_0$). In the
rest frame of matter emerging from the shock (O--A), the velocity at the
head of the simple wave is $v_w''(\epsilon_B)$ as given by (\ref{vw}), and
thus, in the rest frame of incoming matter, $v_B \equiv [v_w''(\epsilon_B)
+v_{CJ})]/[1+v_w''(\epsilon_B)\, v_{CJ}] \simeq 0.820$. Therefore,
the CM velocity required to reach point B is $v_{CM} \equiv v_B$.

Above point B and below C, the compression profile consists of a shock to
the CJ point A, a compressional simple wave, and a second shock.
The position of the first shock and the respective hydrodynamic
quantities are determined as above.
In the rest frame of the second shock, the velocities obey the relations
(\ref{v}), except that quantities with subscript 0 have to be replaced
by the quantities $\tilde{\epsilon},\, \tilde{p}$, and
$\tilde{v}$. In the rest frame of compressed matter, the second shock
moves with velocity
\begin{equation}
v_{sh} \equiv -v = \left[ \frac{(p-\tilde{p})(p+\tilde{\epsilon})}{
(\epsilon-\tilde{\epsilon})(\tilde{p}+\epsilon)} \right]^{1/2}~,
\end{equation}
while matter flows into that shock with velocity
\begin{equation} \label{cond3}
v_w(\tilde{\epsilon}) \equiv \frac{\tilde{v}-v}{1-\tilde{v}\, v}
= - \left[ \frac{(p-\tilde{p})(\epsilon-\tilde{\epsilon})}{
(p+\tilde{\epsilon})(\tilde{p}+\epsilon)} \right]^{1/2}~.
\end{equation}
Using (\ref{vw}) (after boosting it to the global CM frame)
for the left hand side, this equation represents
a condition to determine $\tilde{\epsilon},\, \tilde{p}$. In practice,
one constructs the compressional simple wave starting from the
CJ point as in the previous case,
but simultaneously checks for each $v_w(\epsilon_w)$ whether the
condition (\ref{cond3}) is fulfilled. At first, the left hand side
will be {\em larger\/} than the right due to the following reason:
a compressional wave which is too ``short'' implies a shock that is too strong
(the corresponding final state lies above the true final state on the wave
adiabat), and a stronger shock is more effective in decelerating matter.
For a given $v_{CM}$, a shock that is too strong
will result in a net {\em positive\/} velocity of matter
in the final compressed state, instead of bringing it only to rest, or in other
words, $v_w(\tilde{\epsilon})$ was not yet sufficiently {\em negative\/}
in order that matter is just stopped by the second shock.
Analytic profiles for the hydrodynamic variables are shown in Fig.\
6 for $v_{CM} = 0.825$.

Point C is numerically determined as $p_C = 2.703\, \epsilon_0,\, X_C=
0.195\, X_0$ ($\epsilon_C = 18.944\, \epsilon_0,\, n_C= 10.525\, n_0$).
{}From this point onwards, single shock solutions are again hydrodynamically
stable. Thus, from (\ref{gamma}) we determine the CM velocity
to reach that point as $v_{CM} \equiv v_C \simeq 0.831$.
Profiles for the hydrodynamic variables are presented in Fig.\ 7
for $v_{CM} = 0.9$.

This concludes the discussion of the analytic construction of the
compression wave profiles.
We finally note that in case point A of the Taub (or generalized
shock) adiabat is not a true CJ point but a kink in the adiabat\footnote{This
could happen for particularly ``stiff'' hadron matter equations of state.
Then, this point corresponds to the phase boundary between hadron and
mixed phase matter.}, also a double shock solution
is possible for certain values of $v_{CM}$ \cite{bugaev}. Since this situation
does not occur for the EoS considered here, we do not discuss it in detail.

\section{Numerical results}

In this section we present numerical solutions for
the compression of nuclear matter. The first test is whether the tables
of thermodynamic quantities constructed as described in Section 2 are
sufficiently accurate to reproduce
the {\em analytical\/} solution presented in the preceding section.
To this end, we solved the Taub equation (\ref{Teq}) and
constructed the generalized shock adiabat {\em using linear interpolation
on these tables}. The curves of Fig.\ 3 are satisfactorily reproduced,
even the wave adiabat, the construction of which
involves the simultaneous solution of eqs.\ (\ref{Teq}) and
(\ref{cond2}). The position of the CJ point can be found within 1\% accuracy.

Then we checked whether the analytic shape of a simple compression wave
can be reproduced.
To this end, a discretized form of the thermodynamical identity (\ref{cs2})
using tabulated values for the pressure was employed in the
integrand of eq.\ (\ref{vw}).
Also in this case, agreement was found to be rather good,
confirming that the tabulation of the thermodynamic
quantities is sufficiently accurate.

The SHASTA and relativistic HLLE algorithms were described in detail in
Ref.\ \cite{test1}.
To test their numerical performance in the ``slab-on-slab''
collision problem, we choose the four different
CM velocities $v_{CM} = 0.7,\, 0.8,\, 0.825,$ and $0.9$, for which
the analytical profiles were explicitly constructed as discussed
in Section 3.
The numerical profiles of the CM frame energy density $T^{00}$ (normalized
to the ground state energy density $\epsilon_0$) and
the temperature $T$ (in MeV) are shown in Figs.\ 4 (a--c)
for the relativistic HLLE ($\Delta t / \Delta x \equiv \lambda = 0.99$)
and in Figs.\ 4 (d--f) for the SHASTA ($\lambda = 0.4$) for $v_{CM} = 0.7$.
(The presentation of quantities in terms of the similarity variable
$\zeta$ is advantageous since in this way one can easily monitor
the approach of the numerical to the analytical solution \cite{test1}.)
The resolution of the shock front is rather good ($\sim 4$ grid cells)
for both algorithms already after a few time steps ($\sim 50$).
The SHASTA produces a small overshoot at the shock front.

In Fig.\ 5 we show the corresponding results for $v_{CM} = 0.8$.
The approach to the analytical solution is slow (about 500 time
steps for both algorithms). At early times the compression
configuration rather resembles a shock front that is
broadened by viscosity \cite{ornik}. The SHASTA run features a
non-propagating instability on the shock plateau at
$x \simeq 25$ (as a function of $\zeta =x/t$, such an instability moves
of course to the left in the course of time, cf.\ Fig.\ 5 (e)).
This instability can be removed by decreasing the antidiffusion fluxes,
see Appendix. The HLLE run shows a slowly decaying
overshoot in the energy density around $x=\zeta=0$.
It is a remnant of the first few time steps when
numerical transients cause an overestimate of the shock plateau
(this effect is clearly visible in Figs.\ 4 and 7). Apart from these
phenomena, both algorithms reproduce
the position and shape of the compressional wave
configuration quite well after about 500 to 1000 time steps.

The origin of the instability (Fig.\ 5 (d)) and the overshoot (Fig.\ 5 (a))
and the reason for their persistence are easily understood
considering the fact that the final state corresponds
to TA mixed phase matter where pressure gradients are small (cf.\ Fig.\ 2 (a);
the same holds for temperature gradients, cf.\ Fig.\ 2 (b), therefore,
corresponding phenomena do not occur in the temperature profiles,
Figs.\ 5 (c,f)).
On one hand, this facilitates compression of matter (one has to exert
less force in the compression) and subsequently the creation of local density
maxima. On the other hand, the usual driving force for the expansion of
local density maxima is absent. They can only decay on account of
numerical diffusion. Therefore, the instability produced by the
standard version of the SHASTA (Figs.\ 5 (d,e)) is removed when
the antidiffusion is decreased, since this increases the numerical diffusion.
On the other hand, the numerical diffusion of the relativistic HLLE is just
large enough to slowly damp out the overshoot at $x=\zeta=0$ (Figs.\ 5 (a,b)).
We finally note that in the mixed phase a density increase (along
the simple compression wave) corresponds
to a temperature decrease, cf.\ Fig.\ 5 (b,c,e,f), reflecting
once more the TA nature of this phase.

Fig.\ 6 shows our results for $v_{CM}=0.825$. While the relativistic
HLLE is well able to reproduce the rather complex compressional wave
configuration (modulo unavoidable numerical dissipation), the SHASTA
fails completely: apart from a slowly growing, non-propagating
instability at $x \simeq 10$, a single shock front forms instead
of the configuration consisting of two shocks and a simple wave.
The reason is again that
the numerical dissipation is too small (see also Appendix).
Reducing the {\em anti}diffusion in the SHASTA
therefore considerably improves the reproduction of the
analytical profile, see Appendix for details.

Finally, in Fig.\ 7 the profiles are shown for $v_{CM}=0.9$. Since
the hydrodynamically stable solution is again a single compression shock,
this case is similar to Fig.\ 4, i.e., both algorithms reproduce
the analytical solution rather well after a few time steps.
Comparing Figs.\ 7 (b) and (e), the resolution of the shock front
in the SHASTA appears to be worse than for the
HLLE. However, one has to remember that, due to the smaller
CFL--number $\lambda$ for the SHASTA run, the profiles extend over fewer cells
after the same number of numerical time steps \cite{test1}.
Close inspection reveals that the shock front is resolved over $\sim 3$
grid cells for {\em both\/} algorithms after about 30 time steps.

In conclusion, despite the simplicity of the numerical algorithms
the (partly rather complex) analytical solutions are reproduced
remarkably well with the relativistic HLLE and also with the SHASTA
after decreasing the antidiffusion.
This constitutes an independent ``numerical'' proof
of the correctness of the arguments presented in \cite{bugaev} and
Section 3.
It is remarkable that an usually unwanted feature like numerical diffusion
is now {\em necessary\/} to reproduce the physical solution. The reason is
that in TA matter this solution reacts sensitively to numerical instabilities.
In order to suppress the latter and consequently
make the transport algorithm more robust, one has to
increase the numerical diffusion.
In the case of single compression shocks
the approach to the analytical solution requires about 50 numerical time
steps. However, to reproduce the more complex
compressional wave configurations for TA matter
takes about a factor 10 to 20 longer.
Consequences for the simulation of heavy--ion collisions are pointed out
in the next section, where the compression of finite systems is studied.
We note that for the HLLE runs we have used a constant velocity of sound
$c_s^2=1/3$ in the signal velocity estimates, cf.\ also \cite{test1}.
Using the physical sound velocity has already been shown to produce unphysical
solutions in the expansion of matter into vacuum \cite{test1}. As
demonstrated in the Appendix, such problems occur also in the compression
\footnote{Of course, the choice $c_s^2=1/3$ is only safe as along as the
{\em physical\/} velocity of sound does not {\em exceed\/} this value.
For instance, for {\em pure\/}
hadronic matter described by the $\sigma-\omega$--model the velocity of
sound even approaches the causal limit for large $\epsilon, n$.
Fortunately, for the EoS constructed in Section 2, the phase transition to
the QGP phase (with $c_s^2=1/3$) prevents this, and
no problems were encountered for the test cases studied here.}.

\section{Compression of finite systems}

We consider a one--dimensional collision of two (equal) nuclei with
rest frame radius $R$. The nuclei have the CM velocity $v_{CM}$
and are initialized in the moment of first contact.
The initial condition for this problem is
\begin{eqnarray}
\epsilon(x,0) & = & \left\{ \begin{array}{ll}
        \epsilon_0~, & |x| \leq 2\, R/\gamma_{CM} \\
        0~, & |x| > 2\, R/\gamma_{CM}~~,
                            \end{array} \right. \\
n(x,0) & = & \left\{ \begin{array}{ll}
        n_0~, &  |x| \leq 2\, R/\gamma_{CM} \\
        0~,  & |x| > 2\, R/\gamma_{CM}~~,
                     \end{array} \right.  \\
v(x,0) & = & \left\{ \begin{array}{ll}
                   v_{CM}~, & -2\, R/\gamma_{CM} \leq x \leq 0\\
                   -v_{CM}~, & 0 < x \leq 2\, R/\gamma_{CM} \\
                   0~, & |x| > 2\, R/\gamma_{CM}~~.
                 \end{array}    \right.
\end{eqnarray}
The compression ends when the
compression front has traversed the nucleus. In the CM frame, this
corresponds to a time \cite{bugaev}
\begin{equation}
t_F = \frac{\gamma_{CM} (1 - v_{sh}' v_{CM})}{v_{sh}'}~R~~,
\end{equation}
where $v_{sh}'$ is the velocity of the compression front in the rest frame
of the incoming nucleus as calculated in Section 3,
i.e., it is either the velocity of the single compression shock in the
case $v_{CM} \leq v_{CJ}$ or $v_{CM} \geq v_C$, or that
of the shock to the CJ point for $v_{CJ} < v_{CM} < v_C$.
In the latter case, $v_{sh}' \simeq 0.878$, see Section 3.
In Fig.\ 8 $t_F$ (in units of $R$) is shown as a function of $v_{CM}$.
If $v_{CM}$ is less than the velocity of sound in the ground state
of nuclear matter $c_{s,0} \equiv (K_0/9\mu_0)^{1/2}$ ($\simeq 0.190$ for
the parameters of Section 2), no shock discontinuity can form \cite{LL6}.
The point where the generalized shock adiabat enters the mixed phase
and also point C on this adiabat (cf.\ Fig.\ 3 (b)) are
visible as kinks in $t_F$.

In order for the numerical algorithms to approach the correct analytical
solution for times $t<t_F$, i.e., before expansion sets in,
the grid spacing has to be sufficiently small.
We have seen in the previous section that in the case of single compression
shocks the shock profile is accurately reproduced after about 50 time steps.
Therefore, the grid spacing should be chosen smaller than $\Delta x \simeq
t_F/50\, \lambda$. To give an example, for $t_F \simeq R$,
$\Delta x < 0.02\, R$ for the HLLE ($\lambda = 0.99$) and
$\Delta x < 0.05 \, R$ for the SHASTA ($\lambda = 0.4$).
For a typical nuclear radius of $R \simeq 5$ fm, a feasible grid spacing for
the HLLE would therefore be $\Delta x < 0.1$ fm and
for the SHASTA $\Delta x < 0.25$ fm.
However, for ultrarelativistic collisions where $t_F \sim 0.2\, R$,
$\Delta x$ should at least be a factor of 5 smaller.

We do not show explicit calculations for the case of compression via
single shock discontinuities,
since the result is obvious: after the incident nuclei
are completely stopped by the compressional shock waves, expansion sets in.
The only remaining question is whether the performance of the algorithms
for this {\em expansion\/} of {\em baryon-rich\/} nuclear matter with a {\em
realistic\/} nuclear matter EoS is of similar quality as for the
(net) baryon-free case studied in Ref.\ \cite{test1}. This investigation
is out of the scope of the present work.

For the complex compressional wave patterns of Figs.\ 5, 6
we have to good approximation $t_F \sim 0.57\, R$ for all
cases, cf.\ Fig.\ 8.
We have seen that of the order of 500 time steps
are necessary to approach the analytical profiles. Thus, the grid
spacing should be chosen smaller than
$ 0.001\, R \simeq 0.005$ fm for the HLLE and
smaller than $ 0.003\, R \simeq 0.015$ fm  for the SHASTA. The numerical effort
to calculate a collision on such a fine grid is definitely prohibitive,
especially for multi--dimensional problems. It is therefore of interest
whether the distortion caused by a too coarse grid spacing really
changes physical observables, such as particle spectra,
in the final state.

The calculation of these spectra requires
assumptions about the particle emission process from the fluid and, ultimately,
about the ``freeze--out'' of the fluid. This is out of the scope of
the present work and will be subject of a forthcoming paper \cite{bernard}.
For the moment, we only study the spatial CM energy density
distribution of the {\em fluid\/}, which serves as input for
a ``freeze--out'' calculation.
We compare the time evolution of this quantity for one--dimensional
collisions of finite nuclei at $v_{CM} = 0.8$, calculated with the HLLE
for $\Delta x = 0.01\, R$ and $\Delta x = 0.001\, R$ (Fig.\ 9), and calculated
with the SHASTA for $\Delta x = 0.025\, R$ and $\Delta x = 0.0025\, R$
(Fig.\ 10). For the SHASTA runs we have reduced the antidiffusion fluxes, since
this stabilizes the algorithm and
improves results considerably, cf.\ Appendix. Let us note that
energy, momentum, and baryon number conservation for all runs is
better than $10^{-3}$.

In Fig.\ 9 one observes that for the compressional stage, Figs.\ 9 (a,b),
the finer grid spacing leads, as expected, to a much more accurate description
of the analytical profile, cf.\ Fig.\ 5. However, the final CM energy
density profiles emerging after the expansion stage, Figs.\ 9 (c,d),
are quite similar. Indeed, the run with the coarser grid spacing
even yields a smoother profile and does not show ``terraces'' as
can be seen in (b,d) on the low density tails in either compression or
expansion stage. Furthermore, small--scale numerical instabilities are observed
around the stationary point of the profile due to the exceedingly
long calculation time in (d).
Consequently, for the HLLE the use of a grid spacing which is
sufficiently fine to reproduce the compressional
stage yields no obvious advantage when one is interested in final
state fluid quantities,
but it is even disadvantageous from the point of view of
the required calculational effort (the same physical time corresponds
to 10 times the number of numerical time steps, the smaller grid spacing
requires 10 times more grid cells on the same physical length scale).

{}From the physical point of view we note that the expansion of the system
proceeds similar as in the (net) baryon-free case studied in \cite{test1}:
the final compressed state consists of TA mixed phase matter, and is
therefore consumed by a rarefaction discontinuity. In Figs.\ 9 (c,d) one
clearly observes this shock wave travelling into the
high density zone.
Hadronic matter is expelled from the shock and subsequently expands
via a simple rarefaction wave.

For the SHASTA (Fig.\ 10) the distorting effect of a too coarse
grid spacing is more obvious. Clearly, not only is the reproduction
of the compression profile insufficient, also the rarefaction shock
wave in the expansion stage is no longer discernible. On the other
hand, the run with the finer grid spacing $\Delta x = 0.0025\, R$
produces excellent results, although there exists again a
tendency to produce ``terraces'' on the tail of the
hadronic expansion wave, as for the HLLE run with $\Delta x = 0.001\, R$.

However, as far as the final state profiles and particle spectra at
freeze--out are concerned, all runs would give similar results.
It is therefore questionable whether high precision calculations with
a small grid spacing are really compulsory for the simulation of heavy--ion
collisions. We note that results for the case $v_{CM}=0.825$ are
rather similar, wherefore we do not show them explicitly.

\section{Conclusions}

In this paper we have studied the compression of nuclear matter in
one--dimensional hydrodynamical ``slab-on-slab'' collisions.
First, a nuclear matter EoS was constructed, which describes the hadron matter
phase in terms of an improved version of the original $\sigma-\omega$--model.
For the QGP phase, the MIT bag model EoS was employed, and both phases were
matched via Gibbs' phase equilibrium conditions for a
first order phase transition. In numerical algorithms to solve relativistic
hydrodynamics, this EoS must be used in tabulated form,
since an ``on-line'' determination of the pressure
(which is required in the solution
of the hydrodynamical equations at various intermediate steps,
cf.\ \cite{test1}) would be calculationally prohibitive.
However, the usefulness of the tables for thermodynamic quantities
is not restricted to the hydrodynamical problems studied here,
they can also be applied to future and, most important, more realistic
multi--dimensional calculations.

Subsequently we have constructed the analytical solution to the
one--dimensional compression of nuclear matter for various incident velocities.
Since nuclear matter with a first order phase transition is TA in
certain regions of the independent thermodynamical variables, the
hydrodynamically stable compression solution is no longer a
single shock discontinuity, as for TN matter, but consists in general
of a sequence of shocks and compressional simple waves. We remark that
the constancy of the specific entropy on compressional simple waves
leads to a plateau in the respective excitation function which
might in turn lead to a corresponding plateau in the excitation function
of the pion multiplicity \cite{BGR}. This would be a clear signal
for the phase transition between hadron matter and the QGP.

Then we have investigated the performance of numerical algorithms
for this particular test problem. Both the relativistic
HLLE and the SHASTA algorithm are well able to reproduce the analytical
profiles, provided the numerical diffusion is sufficiently large.
In a sense, this constitutes an independent numerical
``proof'' for the correctness of the theoretical arguments presented in Ref.\
\cite{bugaev} and in Section 3 of this work concerning the
hydrodynamically stable compression solution.
The algorithms require about an order of magnitude longer
to approach the more complicated compressional solution in the region
where nuclear matter is TA than they need for the reproduction
of the compressional shock waves in TN matter.
To account for this, one has to choose
a sufficiently fine grid spacing $\Delta x$
in collisions of finite nuclei at the respective incident energies,
in order that the analytical profile is reproduced before the compression wave
has traversed the nucleus and expansion sets in.
However, we have found that the final profiles for a calculation
employing a too coarse $\Delta x$ do not differ
significantly from those for a finer $\Delta x$,
so that the computational effort can be considerably reduced,
at least as long as one is interested
in final state observables only\footnote{However, the effect will
be negligible even on observables that are most sensitive
to the {\em initial\/} hot stage,
like photons or dileptons \cite{adrian}, since
the {\em average\/} temperatures are also very similar for the various
runs of Figs.\ 9, 10.}.

We anticipate that the algorithms presented in detail in Ref.\ \cite{test1}
and studied for various non--trivial test problems in \cite{test1} and
the present work will find numerous applications in multi--dimensional
hydrodynamical problems. Of special interest is a deeper understanding
of the flow as discovered in BEVALAC experiments a decade ago
\cite{BEV} and just recently confirmed by quantitatively excellent data
from the EOS--collaboration \cite{rai}. Flow was also observed in recent AGS
experiments \cite{Yingchao}. At present it is unclear whether its
magnitude can be completely accounted for in the framework of microscopic
cascade models \cite{ARC} or whether it is consistent only with
a phase transition in the nuclear matter EoS \cite{DHR}.
\newpage
\noindent
{\bf Acknowledgements}
\\ ~~ \\
D.H.R.\ thanks Miklos Gyulassy for his continuous interest and encouragement
that contributed essentially to the completion of this work.

\appendix
\section{Appendix}

In this Appendix we study the effect of modifications of the algorithms
as used in the main part of this work. First, we demonstrate that the
performance of the SHASTA in the reproduction of the complex compressional
wave configurations of Figs.\ 5, 6 is considerably
improved by decreasing the antidiffusion
fluxes. As discussed in the Appendix of \cite{test1}, to this end the first
factor 1/8 in eq.\ (17) of \cite{test1} is replaced by 1/10. The effect
is shown in Fig.\ 11. One observes that the instabilities in Figs.\ 5, 6
(a,b,d,e) are removed due to the larger numerical dissipation
introduced by reducing the antidiffusion fluxes. Contrasting Fig.\ 6,
the wave configuration for $v_{CM}=0.825$ is now in reasonable
agreement with the analytical profile.

Finally, we discuss the effect of using the physical velocity of sound
in the signal velocity estimates for the relativistic HLLE. In Fig.\ 12
we confront the correct solution for (a) $v_{CM} = 0.8$ and (b)
$v_{CM} = 0.825$ obtained with a constant velocity of sound $c_s^2=1/3$ in
the signal velocity estimates (cf.\ eqs.\ (29, 30) of Ref.\ \cite{test1})
with the corresponding results (c,d)
using the physical sound velocity calculated according
to (\ref{cs2}) on the table of pressure values. Comparing (a) and (c)
one observes that instead
of reproducing the simple compression wave the code now
grossly overestimates the strength of the first shock which even
accelerates matter to positive velocities.
Subsequently, a second {\em rarefaction\/} shock brings matter to rest.
The final state has a smaller energy density than the correct
solution.

A comparison of (b) and (d) reveals that the code fails completely to
reproduce the configuration consisting of a compressional simple wave
between two shocks. Instead,
a single compressional shock is formed which does not decay, as it should
for TA matter. The situation is rather similar to what the
standard version of the SHASTA yields for this case (Fig.\ 6 (e)).
A too small diffusion was then identified as the reason for the failure
to produce the analytical result. This is a natural explanation also for
understanding Fig.\ 11 (d): the physical sound velocity in
the mixed phase is small, and a small sound velocity decreases the signal
velocity estimates \cite{test1}, and consequently, the numerical diffusion.

\newpage
\noindent
{\bf Figure Captions:}
\\ ~~ \\
{\bf Fig.\ 1:} Nuclear matter phase diagram in the (a) $T-\mu$, (b)
$T-n/n_0$, (c) $\epsilon/\epsilon_0-T$, and (d) $\epsilon/\epsilon_0-
n/n_0$ plane. The dotted line in (c) corresponds to the $n=0$--axis
in (b), the dotted line in (d) to the $T=0$--axis
in (c).
\\ ~~ \\
{\bf Fig.\ 2:} (a) $p/\epsilon_0$, (b) $T$, (c) $\mu$, and (d) $s/n_0$
as functions of $\epsilon/\epsilon_0$ and $n/n_0$. (For presentation
purposes, only every second meshpoint on the $201 \times 240$ mesh is
plotted.)
\\ ~~ \\
{\bf Fig.\ 3:} (a) Single shock Taub adiabat with the
center $(X_0,p_0)$ (point O) and the CJ point A.
(b) Generalized shock adiabat.
The part (O--A) and that above C are identical to the Taub adiabat
in (a). Part (A--B) is a Poisson adiabat (B is the inflection point
of this adiabat), part (B--C) the wave adiabat (for details
concerning its construction see text).
The dotted line is the unstable part of the Taub adiabat of Fig.\ 3 (a).
\\ ~~ \\
{\bf Fig.\ 4:} ``Slab-on-slab'' collision for $v_{CM}=0.7$ as
calculated with (a--c) the relativistic HLLE algorithm ($\Delta x = 1,\,
\lambda = 0.99$) and (d--f) the SHASTA ($\Delta x = 1,\, \lambda = 0.4$).
(a,d) CM frame energy density $T^{00}$ (normalized to
the ground state energy density $\epsilon_0$) as a function of $x$ for
$10,20,30,...,100$ time steps, (b,e) $T^{00}/\epsilon_0$
and (d,f) temperature $T$ as functions of the similarity variable
$\zeta \equiv x/t$,  for $10$ (stars), $20$
(squares), $30$ (diamonds), $50$ (dotted line), and
$100$ time steps (dashed line) in comparison
to the analytical result (full line).
\\ ~~ \\
{\bf Fig.\ 5:} The same as in Fig.\ 4 for $v_{CM} = 0.8$. In (b,c,e,f)
stars correspond to 50, squares to 100, diamonds to 200, the dotted line
to 500, and the dashed line to 1000 time steps.
\\ ~~ \\
{\bf Fig.\ 6:} The same as in Fig.\ 5 for $v_{CM} = 0.825$.
\\ ~~ \\
{\bf Fig.\ 7:} The same as in Fig.\ 4 for $v_{CM} = 0.9$.
\\ ~~ \\
{\bf Fig.\ 8:} The CM time $t_F$ the compression front needs to reach
the edge of the incoming nucleus as a function of $v_{CM}$.
\\ ~~ \\
{\bf Fig.\ 9:} The CM frame energy density (in units of
$\epsilon_0$) as a function of $x$ (in units of the nuclear radius $R$
in its rest frame) for a one--dimensional collision of finite nuclei
at $v_{CM}=0.8$,
calculated with the relativistic HLLE ($\lambda=0.99$) for (a,c)
$\Delta x = 0.01\, R$ and (b,d) $\Delta x = 0.001\, R$.
(a,b) show profiles at constant CM time in the compression stage
(for (a) the time steps are $0,20,40,...,100$,
for (b) $0,200,400,...,1000$), (c,d) the subsequent expansion
(for (c) the time steps are $120,140,...,240$, for (d) $1200,1400,...,2400$).
For the sake of clarity, profiles are alternatingly shown as full and dotted
lines.
\\ ~~ \\
{\bf Fig.\ 10:} The same as in Fig.\ 9 for the SHASTA ($\lambda=0.4$)
with (a,c) $\Delta x = 0.025\, R$ and (b,d) $\Delta x = 0.0025\, R$.
\\ ~~ \\
{\bf Fig.\ 11:} The effect of reducing the antidiffusion fluxes in
the SHASTA, (a--c) as in Figs.\ 5 (d--f), (d--f) as in Figs.\ 6 (d--f).
\\ ~~ \\
{\bf Fig.\ 12:} The effect of using the physical velocity of sound in
the signal velocity estimates in the relativistic HLLE (c,d)
as compared to the correct solution (a,b) (Figs.\ (a,b) are identical
to Figs. 5, 6 (b)).
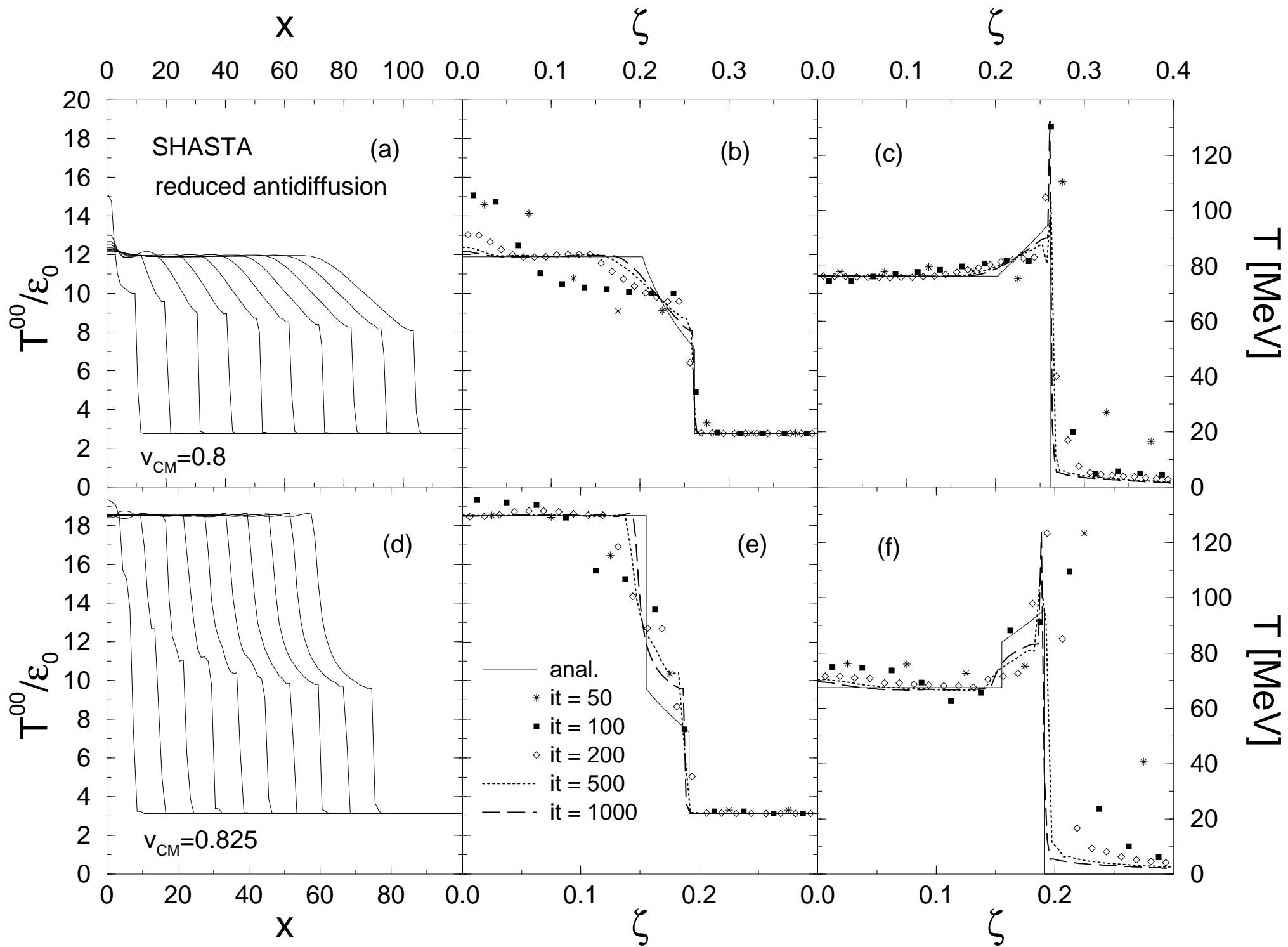
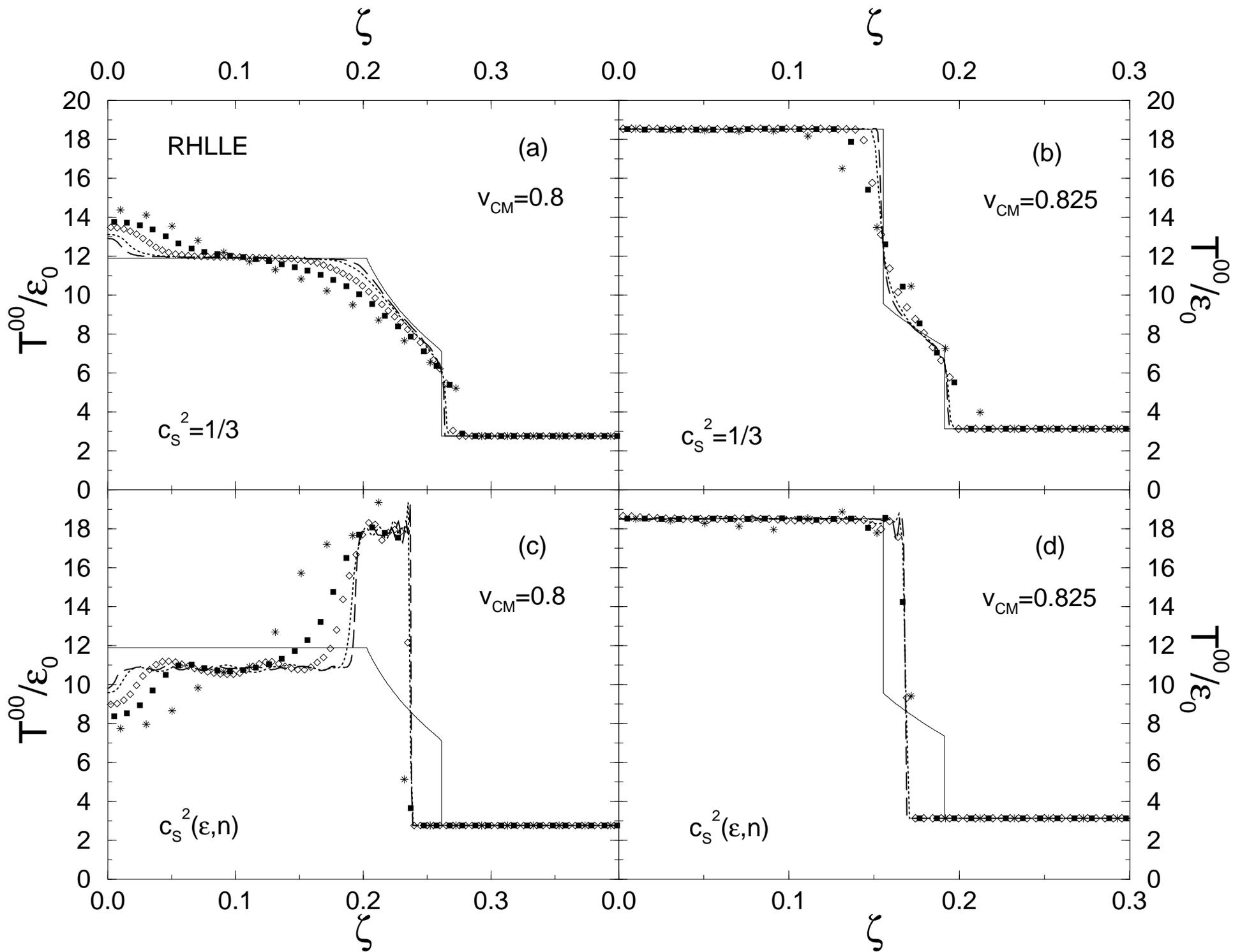
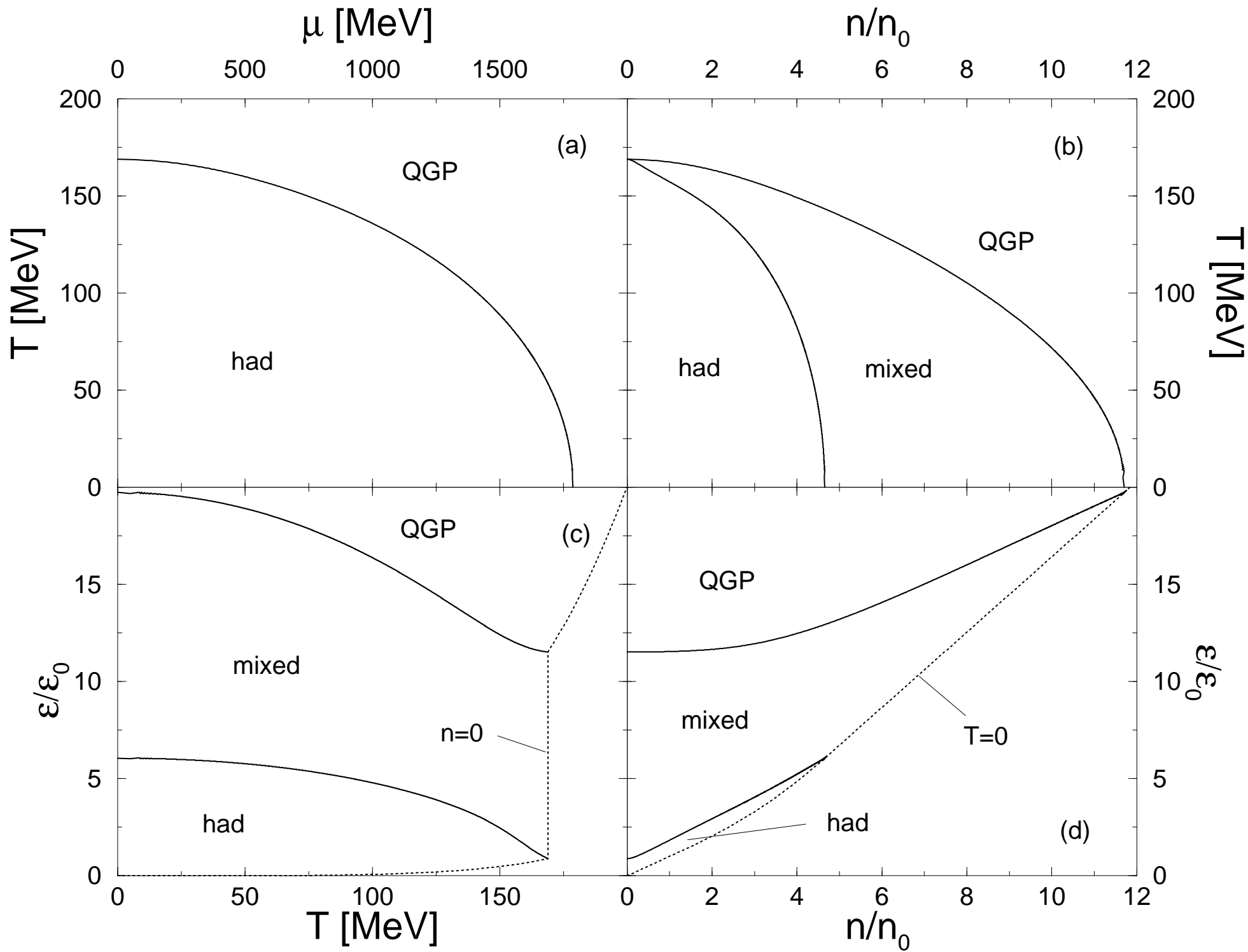
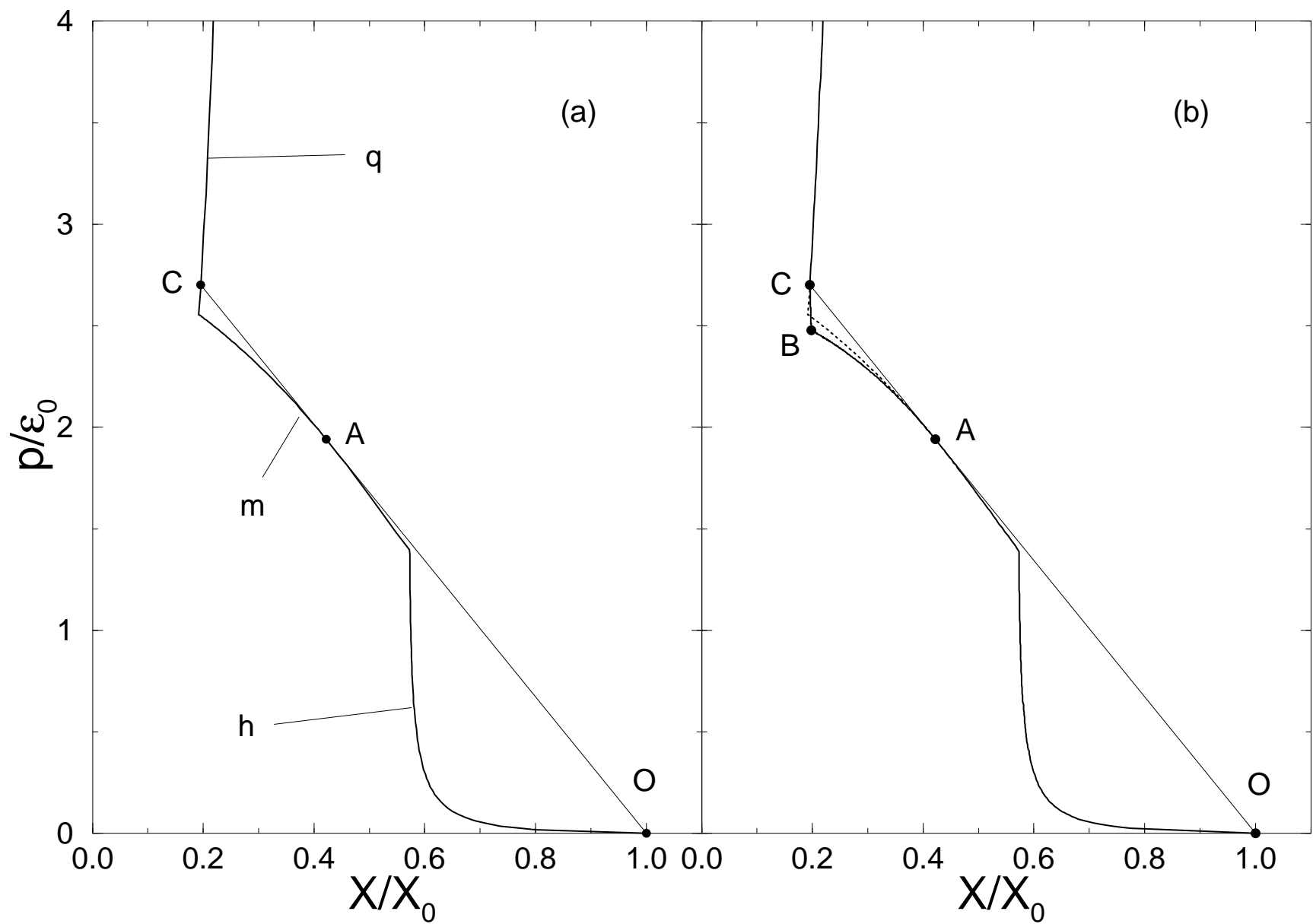
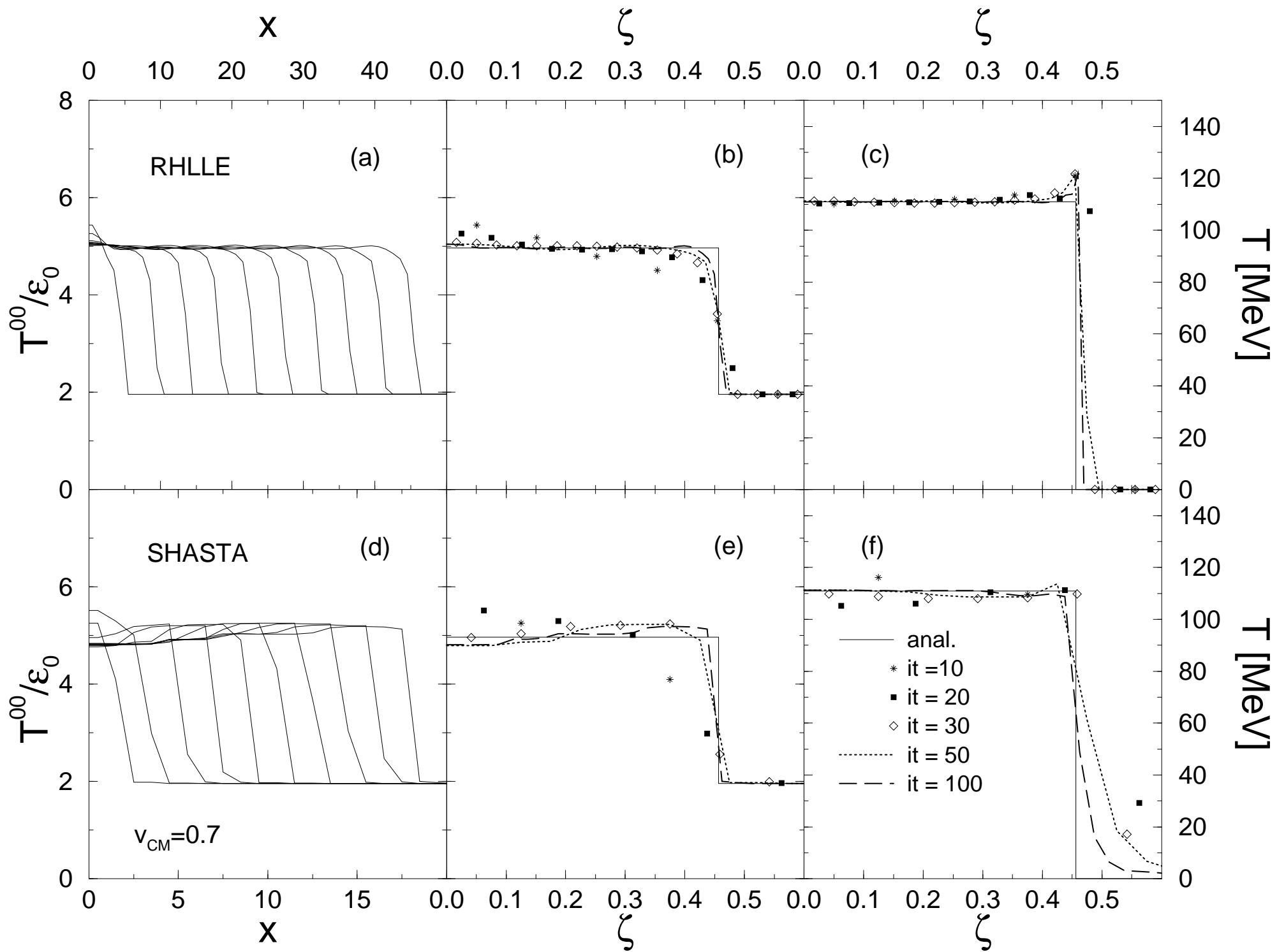
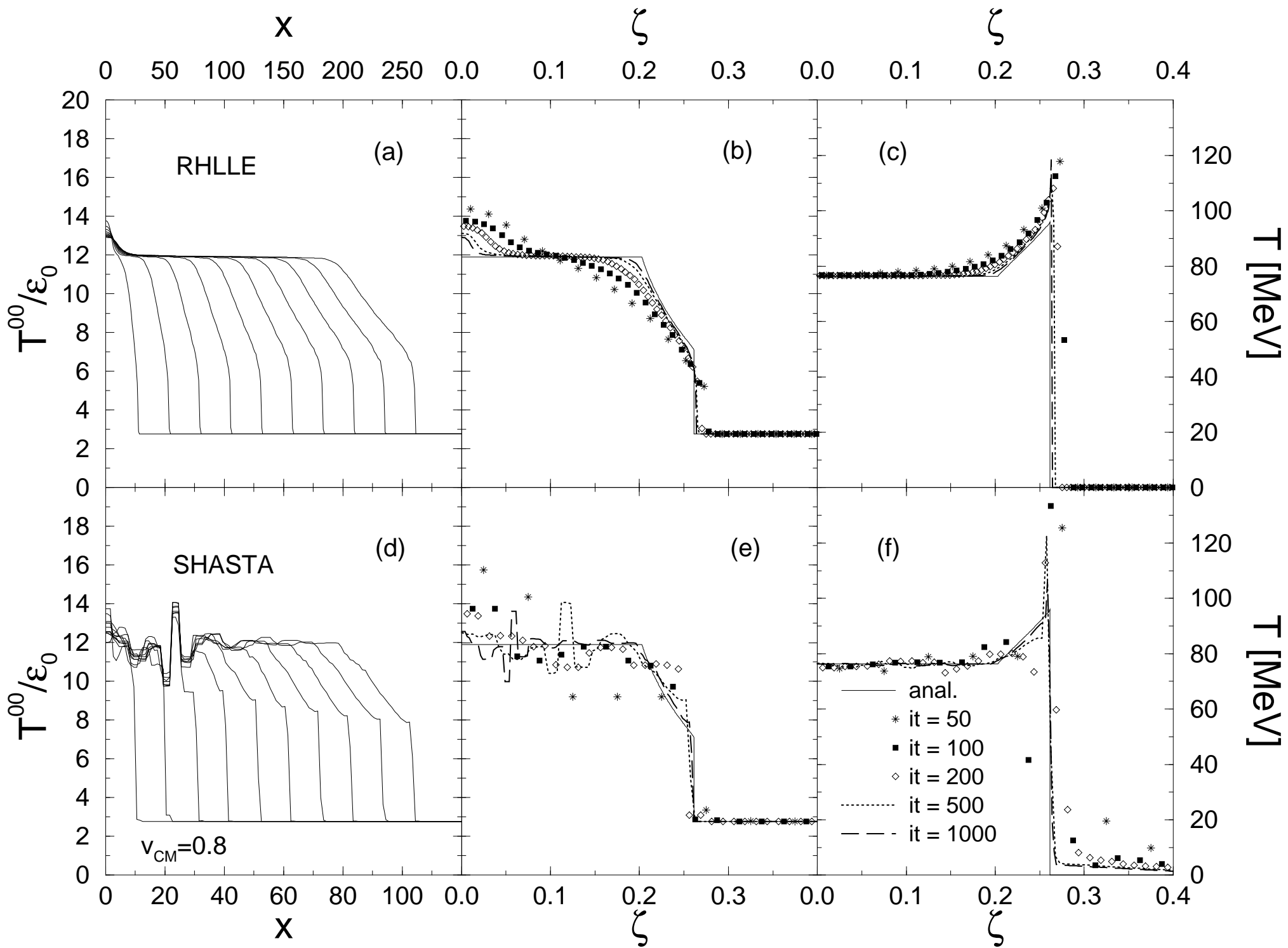
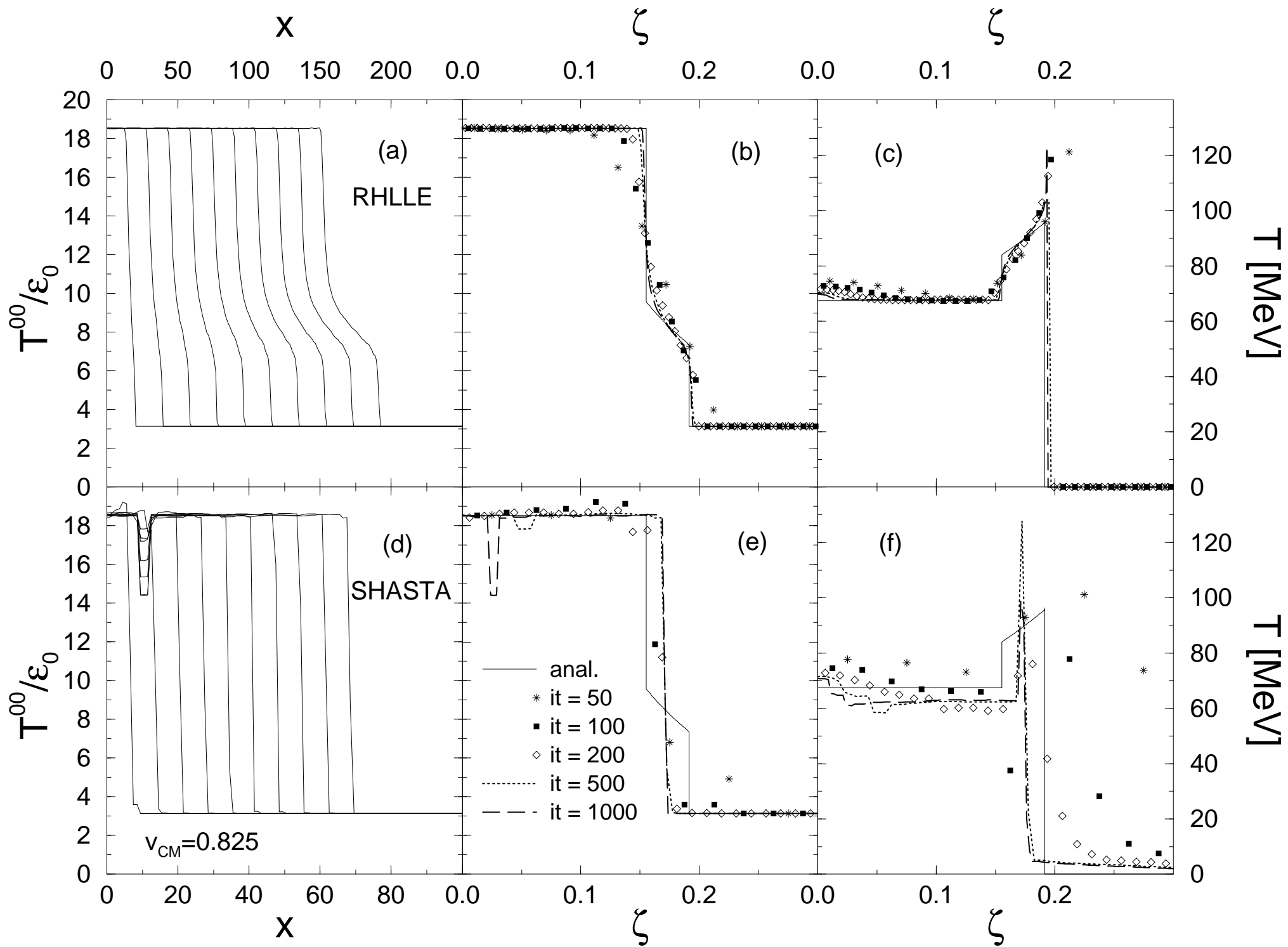
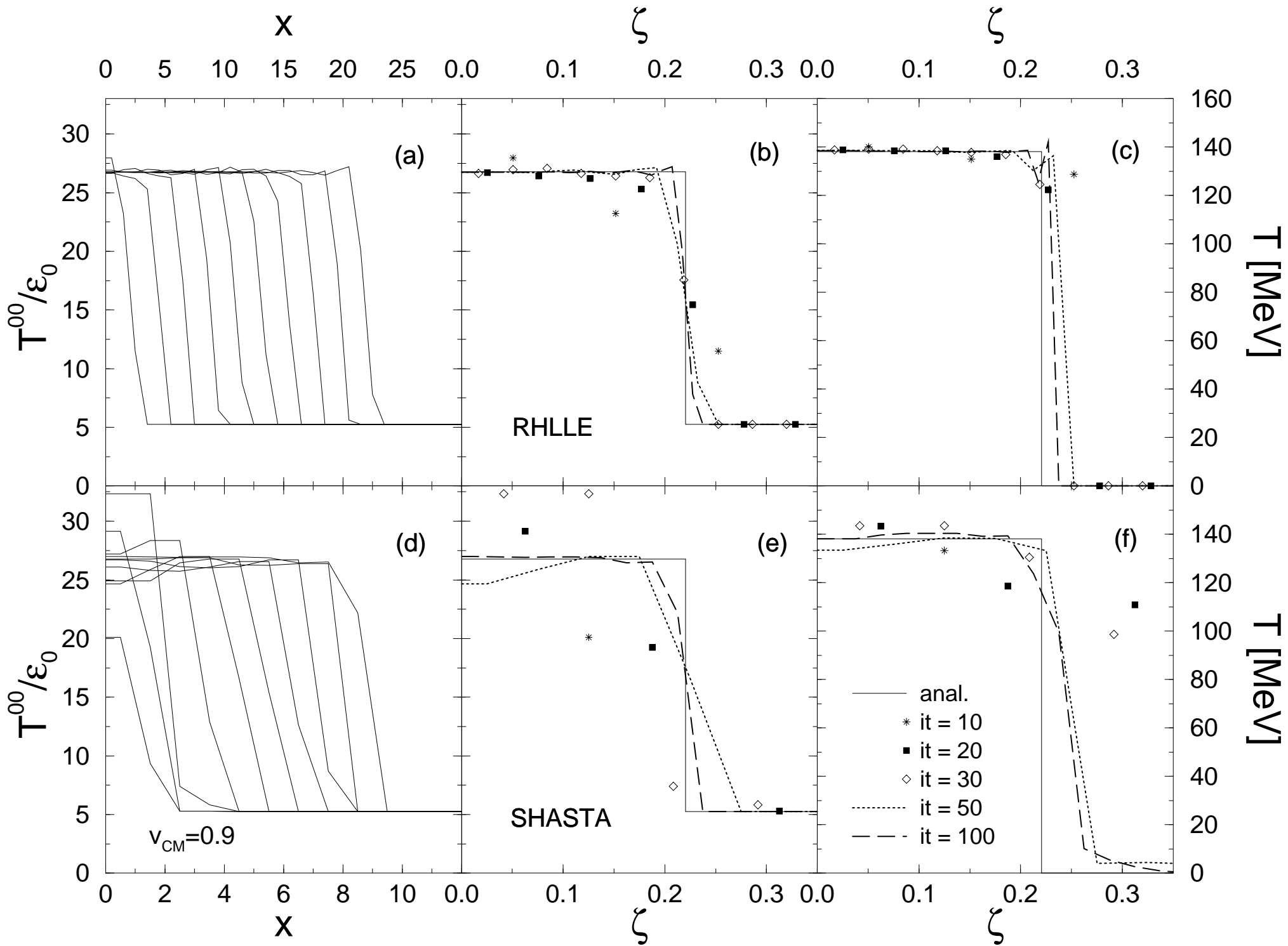
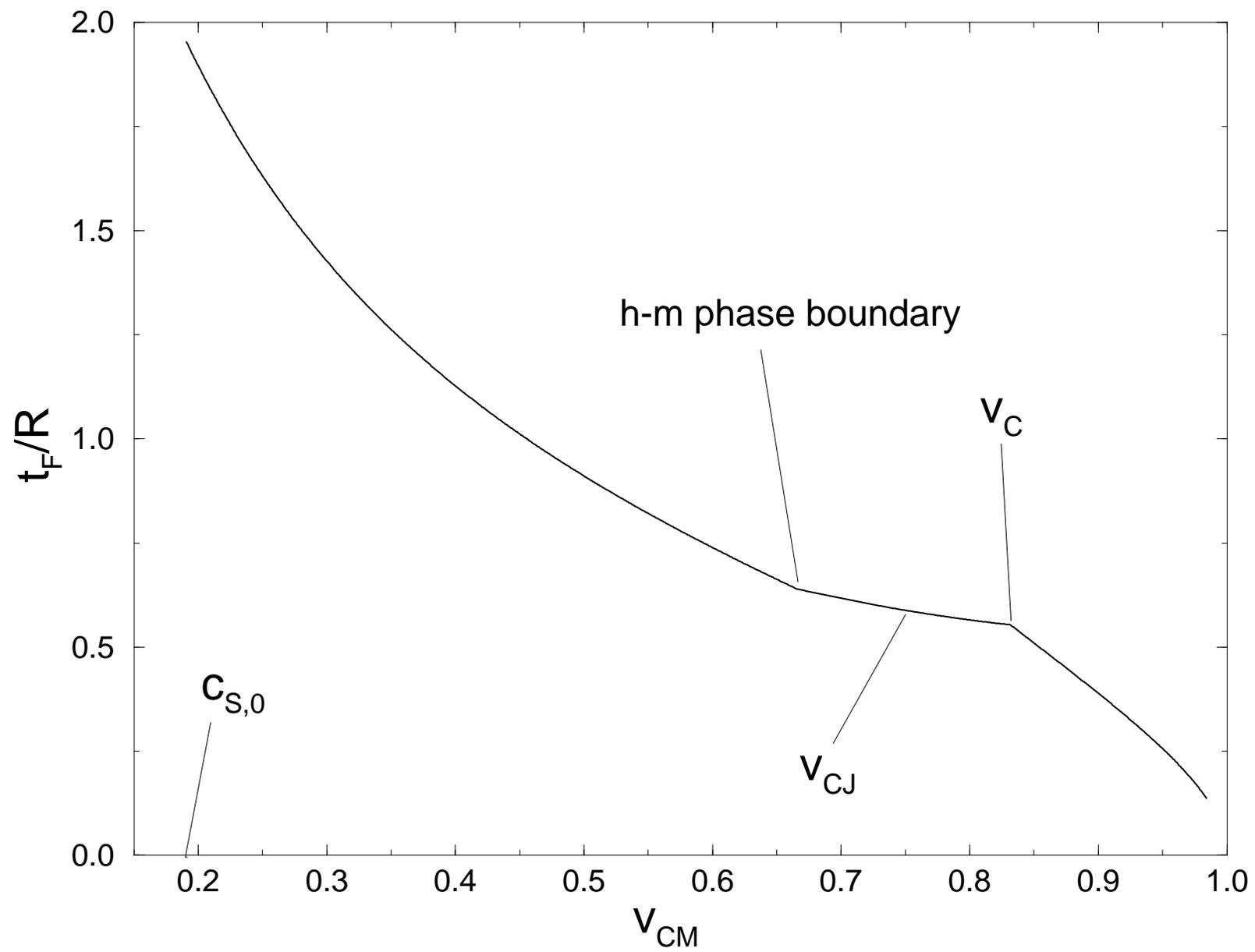
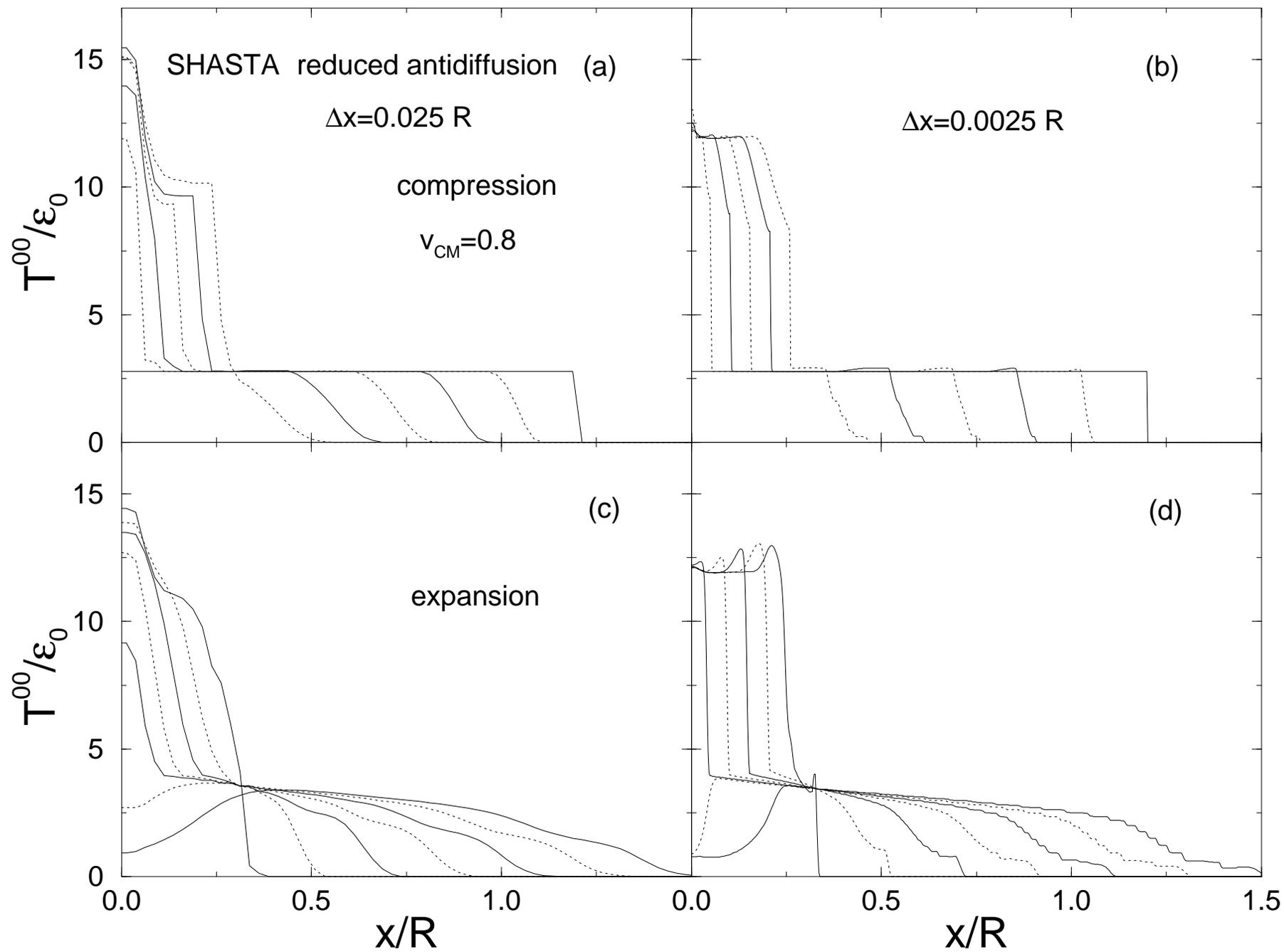
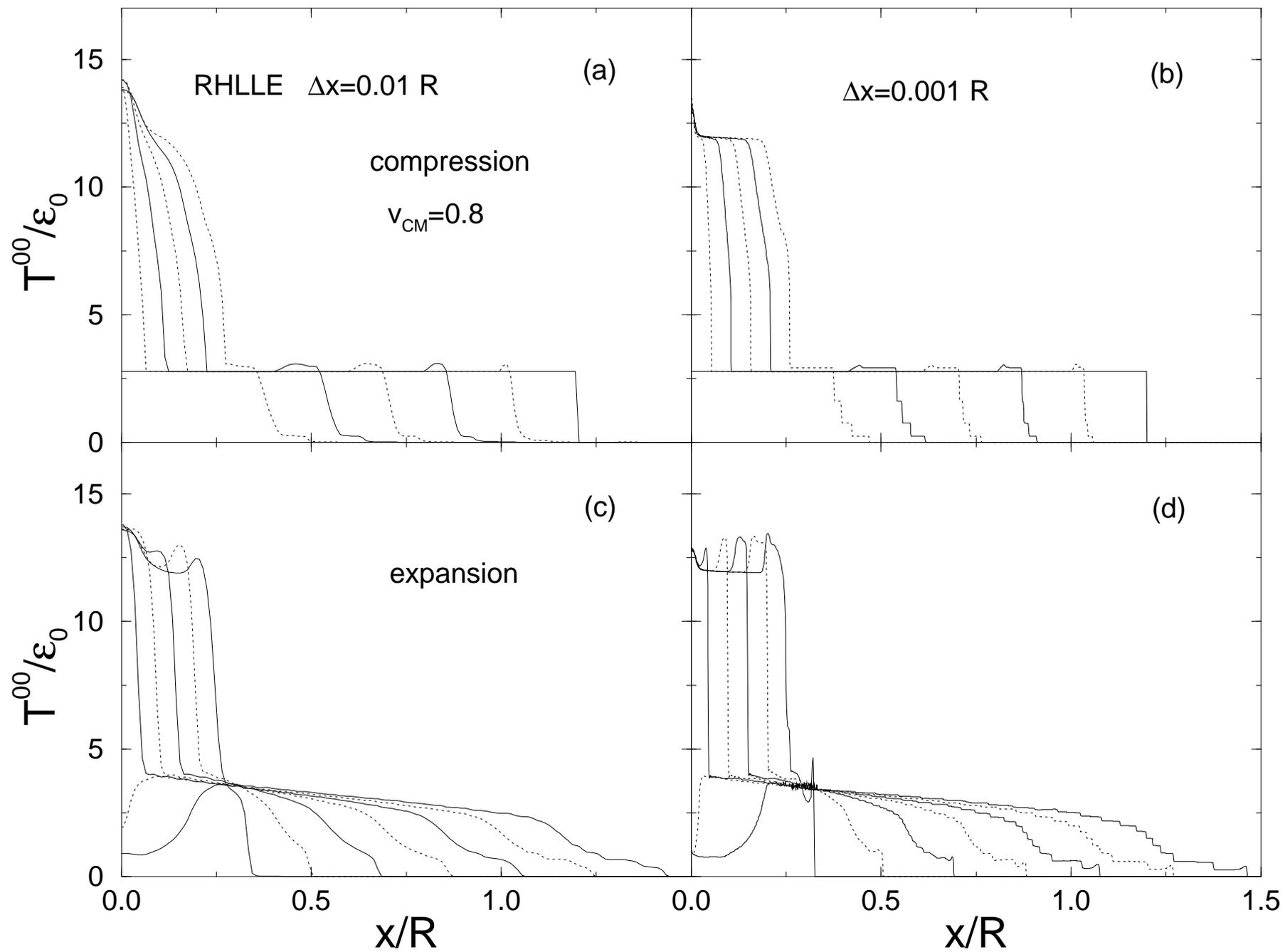
\end{document}